\documentstyle[sprocl]{article}
\input{psfig}
\def\Journal#1#2#3#4{{#1} {\bf #2}, #3 (#4)}
\def\NPA{{\em Nucl. Phys.} A}
\def\NPB{{\em Nucl. Phys.} B}
\def\PLB{{\em Phys. Lett.} B}
\def\PRL{\em Phys. Rev. Lett.}
\def\PRD{{\em Phys. Rev.} D}
\def\ZPA{{\em Z. Phys.} A}
\def\ZPC{{\em Z. Phys.} C}

\def\bea{\begin{eqnarray}}
\def\eea{\end{eqnarray}}

\newcommand{\beq}{\begin{equation}}
\newcommand{\eeq}{\end{equation}}
\newcommand{\la}[1]{\label{#1}}
\newcommand{\ba}{\begin{array}}
\newcommand{\ea}{\end{array}}
\newcommand{\ur}[1]{(\ref{#1})}
\newcommand{\urs}[2]{(\ref{#1},\ref{#2})}
\renewcommand{\vec}[1]{{\bf #1}}

\newcommand{\Eq}[1]{Eq.~(\ref{#1})}
\newcommand{\Eqs}[2]{Eqs.(\ref{#1}, \ref{#2})}

\newcommand{\Tr}{{\rm Tr}\,}

 \def\Sp{\mbox{Sp}}

 \def\Dirac#1{#1\hskip-5.5pt/}
 \def\dd{\Dirac\partial}

 \def\Ug{{U^{\gamma_5}}}

\begin{document}

\title{NUCLEONS AS CHIRAL SOLITONS}
\author{D.I. DIAKONOV}

\address{NORDITA, Blegdamsvej 17, Copenhagen,\\
DK-2100 Denmark}

\address{Petersburg Nuclear Physics Institute, Gatchina,\\
St.Petersburg 188 350, Russia}

\author{V.Yu. PETROV}

\address{Petersburg Nuclear Physics Institute, Gatchina,\\
St.Petersburg 188 350, Russia}

\maketitle\abstracts{
In the limit of large number of colors $N_c$ the nucleon consisting of
$N_c$ quarks is heavy, and one can treat it semiclassically,
like the large-$Z$ Thomas--Fermi atom. The role of the semiclassical
field binding the quarks in the nucleon is played by the pion or
chiral field; its saddle-point distribution inside the nucleon is
called the chiral soliton. The old Skyrme model for this soliton is
an over-simplification. One can do far better by exploiting a
realistic and theoretically-motivated effective chiral lagrangian
presented in this paper. As a result one gets not only the static
characteristics of the nucleon in a fair accordance with the
experiment (such as masses, magnetic moments and formfactors) but also
much more detailed dynamic characteristics like numerous parton
distributions. We review the foundations of the Chiral Quark-Soliton
Model of the nucleon as well as its recent applications to parton
distributions, including the recently introduced `skewed'
distributions, and to the nucleon wave function on the light cone.}

\noindent PACS: 11.15.Pg, 12.38.Aw, 12.39.Fe, 12.39.Ki\\
\noindent Keywords: chiral symmetry breaking, confinement, large $N_c$,
nucleon chiral-quark soliton model, parton distributions

\vspace{1cm}

\tableofcontents

\newpage

\section{Introduction}

Application of the QCD sum rules \cite{SVZ} to nucleons,
pioneered by B.L. Ioffe \cite{Ioffe} has taught us several important
lessons. One is that the physics of nucleons is heavily dominated
by effects of the Spontaneous Chiral Symmetry Breaking (SCSB).
One sees it from the fact that all Ioffe's formulae for nucleon
observables, including the nucleon mass itself, is expressed through
the SCSB order parameter $\langle\bar\psi\psi\rangle$. Therefore,
it is hopeless to build a realistic theory of the nucleon without
taking into due account the SCSB.

The chiral quark--soliton model of the nucleon we are going to review
in this paper is based on two ideas. One is the dominating role the
SCSB plays in the dynamics of the nucleon bound state. The second is,
in fact, a technical tool, not a matter of principle. The idea is that
our world with the number of colors $N_c=3$ is not qualitatively
different from an imaginary world with large number of colors. If one
is able to build a nucleon as a bound state of $N_c$ quarks at large
$N_c$ (keeping $N_c$ as a free algebraic parameter) and if,
furthermore, $1/N_c$ corrections are under control, one can reasonably
expect that the picture one gets putting $N_c=3$ is not too far from
reality. We shall see that actually in some cases the leading-$N_c$
predictions are satisfied in nature to a 1\% accuracy (!), that is far
better than one might expect beforehand from a generic $1/N_c\approx
30\%$ correction.

We start from surveying the background of the model: spontaneous
chiral symmetry breaking and the effective chiral lagrangian
(E$\chi$L) including the Wess--Zumino term. We formulate a simple but
realistic E$\chi$L which is compatible with the phenomenology and
follows from certain theoretical considerations. The properties
of this E$\chi$L is studied in some detail. For that purpose we
introduce a technique of gradient expansion of functional
determinants, which is widely used in many other physical applications.
We then build the semi-classical model of the nucleon, justified at
large $N_c$, and present another technical tool, that of the
quantization of soliton rotations. In the context of the nucleon
problem it is necessary to describe different spin-isospin states of a
nucleon as well as that of the $\Delta$ resonance. Finally, we briefly
describe recent development in the field, namely, the calculation of
various parton distributions inside the nucleon at low virtuality.
Parton distributions obtained in the model satisfy all general
requirements: they are positive, satisfy appropriate sum rules, etc.
In addition, they appear to be in a good accordance with the
data (where measured) without any fitting parameters whatsoever.

\section{How do we know chiral symmetry is spontaneously broken?}

The QCD lagrangian with $N_f$ massless flavors is known to
posses a large global symmetry, namely a symmetry under $U(N_f)\times
U(N_f)$ independent rotations of left- and right-handed quark fields.
This symmetry is called {\em chiral}. [The word was coined by
Lord Kelvin in 1894 to describe molecules not superimposable on its
mirror image.] Instead of rotating separately the 2-component Weyl
spinors corresponding to left- and right-handed components of
quark fields, one can make independent vector and axial $U(N_f)$
rotations of the full 4-component Dirac spinors -- the QCD lagrangian
is invariant under these transformations too.

Meanwhile, axial transformations mix states with different
P-parities.  Were that symmetry exact, one would observe
parity degeneracy of all states with otherwise the same quantum
numbers.  In reality the splittings between states with the same
quantum numbers but opposite parities are huge. For example, the
splitting between the vector $\rho$ and the axial $a_1$ meson is $(1260
- 770)\simeq 500\;{\rm MeV}$; the splitting between the nucleon and its
parity partner is even larger:  $(1535 - 940)\simeq 600\;{\rm MeV}$.

The splittings are too large to be explained by the small bare or
current quark masses which break the chiral symmetry from the
beginning. Indeed, the current masses of light quarks are: $m_u \simeq
4\;{\rm MeV},\;\;m_d\simeq 7\;{\rm MeV},\;\;m_s\simeq 150\;{\rm MeV}$.
The only conclusion one can draw from these numbers is that the chiral
symmetry of the QCD lagrangian is broken down {\em spontaneously}, and
very strongly. Consequently, one should have light (pseudo) Goldstone
pseudoscalar hadrons -- their role is played by pions which indeed are
by far the lightest hadrons.

The order parameter associated with chiral symmetry breaking is
the so-called {\em chiral} or {\em quark condensate}:

\beq
\langle\bar\psi\psi\rangle\simeq -(250\;{\rm MeV})^3
\la{chcond}\eeq
at the scale of a few hundred MeV.
It should be noted that this quantity is well defined only for
massless quarks, otherwise it is somewhat ambiguous. By definition,
this is the quark Green function taken at one point; in momentum space
it is a closed quark loop. If the quark propagator has only the `slash'
term, the trace over the spinor indices understood in this loop gives
an identical zero. Therefore, chiral symmetry breaking implies
that a massless (or nearly massless) quark develops a non-zero
dynamical mass (i.e. a `non-slash' term in the propagator). There are
no reasons for this quantity to be a constant independent of the
momentum; moreover, we understand that it should anyhow vanish at large
momentum.  The value of the dynamical mass at small virtuality can be
estimated as one half of the $\rho$ meson mass or one third of the
nucleon mass, that is about

\beq
M(0)\approx 350\;{\rm MeV};
\la{M0}\eeq
this quantity is directly related to chiral symmetry breaking and should
emerge together with the condensate \ur{chcond} in any theoretical
description of the SCSB.

\section{Effective Chiral Lagrangian (E$\chi$L)}

Let us consider QCD in the chiral limit, meaning that we put the light
current quark masses to zero. For $u,d$ quarks this is a good
approximation to reality: it is known that most physical observables
computed in the chiral limit differ from their true values by about 5\%,
unless the quantity is for some special reasons divergent in the chiral
limit.  For example, the nucleon mass in the chiral limit is lighter by
about 5\%: this is the contribution of the $\sigma$-term to the nucleon
mass. For the $s$ quark this idealization has worse accuracy, about
20\%. We would like to keep the number of light flavors $N_f$ a free
parameter putting it equal to 3 or 2 depending on whether we
incorporate strangeness or not.

Once chiral symmetry is spontaneously broken, the lightest degrees of
freedom in QCD are $N_f^2-1$ pseudoscalar Goldstone bosons (we shall
call them `pions' for short even in the case of $N_f=3$ where four kaons
and the $\eta$ meson are also Goldstone bosons). Therefore, in the
low-momenta region QCD is reduced to a theory of massless interacting
Goldstone bosons. The appropriate nonlinear lagrangian describing
Goldstone pions is called the effective chiral lagrangian (E$\chi$L).
It is convenient to write the pion fields in terms of a unitary
$(N_f^2-1)\times(N_f^2-1)$ matrix,

\beq
U(x)=\exp\,i\,\frac{\lambda^A\,\pi^A(x)}{F_\pi},
\la{U}\eeq
where $\lambda^A$ are eight Gell-Mann matrices in case we consider
$N_f=3$ flavors and three Pauli matrices for $N_f=2$. We divide
the exponent by a pion decay constant $F_\pi=94\;{\rm MeV}$ in order to
use a properly normalized pion field with a standard kinetic
energy, see below.

\subsection{General properties}

In the chiral limit pions are strictly massless, so the E$\chi$L needs
to have at least two derivatives of the matrix $U$, or more. Another
general requirement for building the E$\chi$L is that it
should be chiral-invariant, meaning invariance under a global
$SU_L(N_f)\times SU_R(N_f)$ rotation,

\beq
U\rightarrow AUB^\dagger,\qquad U^\dagger\rightarrow
BU^\dagger A^\dagger,
\la{glob}\eeq
with constant $SU(N_f)$ matrices $A,B$. It is useful to introduce
hermitian matrices

\beq
L_\mu=iU\partial_\mu U^\dagger=\frac{1}{F_\pi}
\partial_\mu\pi^A\,\lambda^A+\ldots
\la{L}\eeq
transforming under the rotation \ur{glob} as $L_\mu\to AL_\mu
A^\dagger$.

A general form of the E$\chi$L can be presented as a series in the
number of derivatives of the fields; it is called the {\em derivative
expansion}. The leading term is the two-derivative one; there is only
one possibility to write it, compatible with the invariance \ur{glob}:
\bea
\la{L2}
{\cal L}^{(2)}&=&\frac{F_\pi^2}{4}\;\Tr L_\mu L_\mu=
\frac{F_\pi^2}{4}\;\Tr \partial_\mu U^\dagger\partial_\mu U\\
\nonumber
&=&\frac{1}{2}\left(\partial_\mu\pi^A\right)^2
+\frac{1}{3F_\pi^2}\,f^{ABE}f^{CDE}\,\partial_\mu\pi^A\pi^B
\partial_\mu\pi^C\pi^D + O(\pi^6).
\eea
The first term is the usual kinetic energy for pions, while the second
(quartic term) describes the $p$-wave pion scattering.  The Noether
current associated with the left-hand transformation in \ur{L2} by an
$x$-dependent matrix $A$ is

\beq
J_{L\mu}^A=-\frac{F_\pi^2}{2}\,\Tr\left(L_\mu\lambda^A\right)=
-F_\pi\partial_\mu\pi^A+\ldots
\la{curr}\eeq
Charged pions decay owing to this current which couples to the $W^\pm$
bosons, hence the normalization factor $F_\pi$ in the above equations.

In the four-derivative order there are several invariants,

\bea
\nonumber
{\cal L}^{(4)}&=&L_1\Tr(L_\mu L_\mu)\Tr(L_\nu L_\nu)
+L_2\Tr(L_\mu L_\nu)\Tr(L_\mu L_\nu)\\
\nonumber
\\
\la{L4}
&+&L_3\Tr(L_\mu L_\mu L_\nu L_\nu)
+C_4\Tr(\partial_\mu L_\mu)^2
\eea
where we indroduce the coefficients $L_1,\ldots$ in accordance with
notations of Gasser and Leutwyler.\cite{GL} These coefficients have
been fitted to the $d$-wave pion scattering and other low-energy data,
and their present-day values are:$\,$\cite{BCG}

\beq
L_1\approx 2L_2,\qquad L_2=(1.35\pm 0.3)10^{-3},\qquad
L_3=(-3.5\pm 1.1)10^{-3}.
\la{LL}\eeq
The equation $L_1=2L_2$ is a prediction of large $N_c$; the term
with $C_4$ is usually expelled from the four-derivative family as
it affects the physical amplitudes only at higher orders; at $N_f=2$
the three first terms in \Eq{L4} are not independent. For these reasons
the four-derivative terms of the E$\chi$L are sometimes presented
as

\bea
\nonumber
{\cal L}^{(4)}&=&(2L_2+L_3)\Tr(L_\mu L_\mu L_\nu L_\nu)
+L_2\Tr(L_\mu L_\nu L_\mu L_\nu)\\
\la{L41}
&=&(3L_2+L_3)\Tr(L_\mu^2L_\nu^2)+\frac{L_2}{2}\Tr\left[L_\mu
L_\nu\right]^2
\eea
where $2L_2+L_3$ is compatible with zero; the last term is called the
Skyrme term.

In the six-derivative order there are rather many invariants but only
limited information is known about them from
phenomenology.$\,$\cite{Knecht}

\subsection{Wess--Zumino term}

There is another famous term in the four-derivative order, called the
Wess--Zumino term.$\,$\cite{WZ} It has several interesting peculiarities.
First, it cannot be written down in a local form as \Eqs{L2}{L4}. One
of the forms has been suggested by Eides and one of the authors
\cite{DE} as an integral over a parameter:

\bea
\la{WZDE}
{\cal L}^{WZ}&=&\frac{N_c}{24\pi^2}
\int_0^1 ds\,\epsilon_{\alpha\beta\gamma\delta}
\,\Tr\Bigl(e^{-is\Pi}\partial_\alpha e^{is\Pi}\Bigr)\Bigl(_\beta\Bigr)
\Bigl(_\gamma\Bigr)\Bigl(_\delta\Bigr) \\
\nonumber
&=&\frac{2N_c}{15\pi^2}\;\Tr\Bigl(\Pi\,\partial_\alpha\Pi\,
\partial_\beta\Pi\,\partial_\gamma\Pi\,\partial_\delta\Pi\Bigr)
+O(\pi^7),\qquad \Pi=\pi^A\lambda^A/F_\pi.
\eea
This expression iz nonzero only starting from
$N_f\geq 3$. The leading $O(\pi^5)$ term gives the low-energy
amplitude of the process $\bar KK\to\pi\pi\pi$.
For two flavors the Wess--Zumino term is zero, at least perturbatively,
in the sense that all terms of the expansion in powers of the pion
field are zero. However, as shown by Witten,$\,$\cite{W} for certain
`large' fields it is nonzero, even for $U\in SU(2)$. Also, the
variational derivatives of \Eq{WZDE} are nonzero. For example, one may
wish to `gauge' the Wess--Zumino action by replacing the derivatives in
\Eq{WZDE} by covariant derivatives of the flavor $SU_L(N_f)\times
SU_R(N_f)$ group. This is necessary if one likes to learn how pions
couple to the photon and the $W,Z$ bosons. The task of gauging
\Eq{WZDE} is not a trivial one \cite{W} given the nonlocal form of the
Wess--Zumino action. The complete and correct result has been given by
Dhar, Shankar and Wadia.$\,$\cite{Dhar} The gauged Wess--Zumino action
is nonzero for any $N_f$ including $N_f=2$. It gives, in particular, the
low-momenta limit of the processes $\pi^0\to \gamma\gamma$ and
$\gamma\to \pi\pi\pi$. In a more common language these processes are
determined by the axial anomaly. Therefore, the Wess--Zumino term in
the E$\chi$L incorporates the axial anomaly, but in a $SU_L(N_f)\times
SU_R(N_f)$ invariant way.

Last, let us present the Wess--Zumino term in a 5-dim form suggested by
Witten.$\,$\cite{W} The action corresponding to the lagrangian \ur{WZDE}
can be written as

\beq
S^{WZ}[\pi]
=-\frac{N_c}{240\pi^2}
\int\!d^5x\,\epsilon_{\alpha\beta\gamma\delta\epsilon}\,
\Tr\Bigl(L_\alpha L_\beta L_\gamma L_\delta L_\epsilon\Bigr).
\la{WZW}\eeq
In this equation it is understood that the pion field $U$ is
continued without singularities from the physical 4-dim space-time to a
fifth dimension such that the physical space-time serves as a border of
the 5-dim `disk'.  Actually, it can be shown that the integrand in
\Eq{WZW} is a full derivative, so that \Eq{WZW} does not depend on the
concrete way one continues the pion field inside the `disk'. The above
\Eq{WZDE} can be viewed as a specific continuation procedure, with the
parameter $s$ playing the role of the `disk' radius.

If the light current quark masses are nonzero, the pion fields become
massive (they are called pseudo-Goldstone fields in this case), and
there appear terms in the E$\chi$L without derivatives,

\beq
\Tr\, m(U+U^\dagger),\qquad \Tr\,mUmU^\dagger, \qquad
\Tr m(U+U^\dagger)L_\mu L_\mu,\qquad{\rm etc.}
\la{Lm}\eeq
where $m$ is the matrix of current quark masses. However, in this
paper we shall restrict ourselves to the chiral limit and neglect such
terms.

Were QCD exactly solvable, one would be able to derive all terms of the
derivative expansion of the E$\chi$L. The dimensionfull coefficients
(like $F_\pi$, etc.) should be expressible through the only
dimensional constant one has in the chiral limit of QCD, namely
$\Lambda_{{\rm QCD}}$, this quantity appearing, via the transmutation
of dimensions, as a renormalization-invariant combination of the
ultra-violet cutoff $\mu$ and the gauge coupling given at this cutoff,
\bea
\la{LambdaQCD}
\Lambda_{{\rm QCD}}
&=&\mu\left[\frac{16\pi^2}{bg^2(\mu)}\right]^{\frac{b^\prime}{2b^2}}
\exp\left[-\frac{8\pi^2}{bg^2(\mu)}\right]
\left[1+O\left(\frac{1}{g^2(\mu)}\right)\right],\\
\nonumber
b&=&\frac{11}{3}N_c-\frac{2}{3}N_f,\qquad
b^\prime=\frac{34}{3}N_c^2-\frac{13}{3}N_cN_f+\frac{N_f}{N_c}.
\eea
So far QCD is not exactly solved, even in the large $N_c$ limit,
therefore, to build a realistic E$\chi$L one has to rely upon models,
physical considerations and comparison with phenomenology.

\subsection{Skyrme model}

A very simple model is named after Skyrme.\cite{Skyrme} It consists
in truncating the infinite series of the derivative expansion at the
four-derivative level, moreover, leaving only one term in \Eq{L4}, the
`Skyrme term',

\beq
{\cal L}^{{\rm Skyrme}}=\frac{F_\pi^2}{4}\Tr L_\mu L_\mu
+\frac{1}{32e^2}\Tr[L_\mu L_\nu]^2.
\la{Skyrme}\eeq
Skyrme has suggested this simplified version of the chiral lagrangian
to find a static configuration of the pion field
minimizing the appropriate energy functional. He conjectured that
this static configuration should be associated with the nucleon.
More than two decades later Witten \cite{W} has revitalized the
original Skyrme's idea showing that, if supplemented by the
Wess--Zumino term \ur{WZW}, a static local minimum of the E$\chi$L can
be, indeed, associated with a nucleon in the large-$N_c$ limit of QCD.
The model of a nucleon based on the local minimum of \Eq{Skyrme} has
been presented by Adkins, Nappi and Witten \cite{ANW} and called the
{\em skyrmion}.

Simple as it is, the model cannot be fully realistic because we know
that there are other terms with four derivatives, numerically of the
same magnitude as the commutator term in \ur{Skyrme}, not to mention
higher-derivative terms. Adding the Wess--Zumino term simply `by hands'
looks somewhat {\it ad hoc}, as is the choice of the simplified
E$\chi$L.

\subsection{Low-momenta limit of QCD}

It has been first noticed in ref. \cite{DE} that the Wess--Zumino term
can be obtained as a functional integral over the quark fields. We
shall show it later in section 4 and now let us consider the
interaction of quarks with the chiral field. As explained in section 2,
an inevitable consequence of SCSB is the appearance of a dynamical or
constituent quark mass $M$, a `non-slash' term in its propagator. A
free lagrangian,

\beq
{\cal L}^{{\rm free}}=\bar\psi(i\dd-M)\psi,
\la{Lfree}\eeq
is, however, not invariant inder axial rotations,

\beq
\psi\rightarrow\exp(i\alpha^A\lambda^A\gamma_5)\psi,\qquad
\bar\psi\rightarrow\bar\psi\exp(i\alpha^A\lambda^A\gamma_5),
\la{axrot}\eeq
because of the mass term.
To construct a lagrangian which is invariant under axial rotations
Goldstone pion fields need to be involved. The simplest lagrangian
invariant under \ur{axrot} is
\bea
\la{chirla}
{\cal L} &=& \bar\psi(i\dd-M\Ug)\psi,\\
\la{Ug}
\Ug &\equiv & \exp\,i\,\frac{\pi^A\lambda^A\gamma_5}{F_\pi}
=U\frac{1+\gamma_5}{2}+U^\dagger\frac{1-\gamma_5}{2},
\eea
since the rotation of the quark fields here can be
compensated by renaming of pion fields, $U\to AUA$ where
$A=\exp(i\alpha^A\lambda^A)$.

The dynamical quark mass $M$ is not a constant: we know that at large
space-like quark virtualities $M(k)$ has to vanish because of
asymptotic freedom, to mention one reason. Therefore, the interaction
of quarks with chiral fields is necessarily momentum-dependent, i.e.
non-local. To avoid difficult questions of what is going on at
time-like virtualities we prefer to formulate the theory first in the
space-like region, i.e. in Euclidean space. In passing from Minkowski
to Euclidean space we make the substitutions:
\bea
\nonumber
i \int d^4x_M = \int d^4x_E,\qquad i\bar\psi_M &=& \psi^\dagger_E,
\qquad p_M^2=-p_E^2,\\
\nonumber
\gamma_{0M}=\gamma_{4E},\qquad \gamma_{iM} &=& i\gamma_{iE},\qquad
\gamma_{5M}=\gamma_{5E}, \\
\la{MtoE}
{\cal L}_M=\bar\psi(i\dd-M\Ug)\psi &\rightarrow&
{\cal L}_E=\psi^\dagger(i\dd+iM\Ug)\psi.
\eea

We write the partition function to which the full QCD partition
function is reduced at low momenta as a functional integral
over both quark and pion fields: \cite{DP3,DP4}
\bea
\nonumber
{\cal Z}&=&\int\!D\pi^A\int\!D\psi^\dagger D\psi\,
\exp\int\!d^4x \left[\psi^\dagger(x)\, i\dd\, \psi(x)\right.\\
\la{Znash}
&+&\!\!i\!\int\!\frac{d^4k_1d^4k_2}{(2\pi)^8}\left.e^{i(k_1-k_2,x)}\,
\psi^\dagger(k_1)\sqrt{M(k_1)}\,\Ug(x)\sqrt{M(k_2)}\,
\psi(k_2)\right]\!\! .
\eea
\Eq{Znash} shows quarks interacting with chiral fields $U(x)$, with
formfactor functions equal to the square root of the dynamical quark
mass attributed to each vertex where $U(x)$ applies. $\Ug$ is a
$N_f\times N_f$ matrix in flavor and a $4\times 4$ matrix in Dirac
indices; it is a unity $N_c\times N_c$ matrix in color.

\subsection{Digression: chiral symmetry breaking by instantons}

The partition function \ur{Znash} has been actually derived in
Ref.(12,13) from considering light quarks in the QCD instanton
vacuum.$\,$\cite{DP1} Instantons -- large topological fluctuations of
the gluon field -- provide a neat and phenomenologically successful
microscopic mechanism of SCSB. \cite{DP2} Because of the famous
't~Hooft zero modes \cite{tH} the random instanton ensemble can be
viewed as a collection of `impurities' each binding a quark at exactly
zero `energy'. Owing to the quantum-mechanical overlap of quark
wave functions quarks can `hop' from one instanton to another, and get
delocalized. The density of quarks at zero `energy' becomes finite;
this density is exactly the chiral condensate
$\langle\bar\psi\psi\rangle$ via the Banks--Casher formula.$\,$\cite{BC}
In more physical terms, when `hopping' from instanton to anti-instanton
and so forth, quarks change their helicity or chirality; delocalization
means that an infinite number of such flippings are involved, and that
generates the `non-slash' term in the quark propagator, i.e. the
dynamical mass $M(k)$. This quantity is directly related to the
Fourier transform of the 't Hooft zero mode in the field of an
instanton. One can evaluate basic quantities related to the SCSB,
namely $\langle\bar\psi\psi\rangle$, $M(k)$ and $F_\pi$ through the
characteristics of the instanton ensemble, that is their density and
average size, these quantities, in their turn, being related to
$\Lambda_{{\rm QCD}}$ \ur{LambdaQCD} from a variational
calculation.$\,$\cite{DP1,DPW}

Furthermore, instantons induce (non-local) quark interactions;
their boson\-ization leads, at low momenta, to the above partition
function \ur{Znash},$\,$\cite{DP4} for a review see Ref.(19).
However, \Eq{Znash} is probably of a more general nature. For example,
a similar partition function (but without formfactors) follows from
the Nambu--Jona-Lasinio model.$\,$\cite{NJL} However, contrary to the
NJL model, instanton-induced interaction preserve all the symmetries
of the original QCD: they are $SU_L(N_f)\times SU_R(N_f)$ invariant
(in case of $N_c=N_f=2$ they are $SU(4)$ invariant) but break
explicitly the axial $U_A(1)$ invariance, which is needed to solve
the $U(1)$ problem.$\,$\cite{DP4,D4}

\subsection{Local field theory at low momenta}

However realistic is the instanton mechanism of SCSB, actually the
low-energy partition function \ur{Znash} is of a very general nature
and in principle one needs not refer to instantons to use it. To make it
even more general one can add to \Eq{Znash} a $Z(k)$ factor in the
kinetic-energy term for quarks and an additional formfactor function
for pions in the interaction term. However, such modifications can be
removed by a redefinition of the $\psi$ and $U$ fields.

The formfactor functions $\surd M(k)$ for each quark line attached to
the chiral vertex automatically cut off momenta at some characteristic
scale. In the instanton derivation of \Eq{Znash} this scale is the
inverse average size of instantons, $1/\bar\rho\approx 600\;{\rm MeV}$.
In the range of quark momenta $k\ll 1/\bar\rho$ one can neglect this
non-locality, and the partition function \ur{Znash} is simplified to a
local field theory:
\beq
{\cal Z}=\int\!D\pi^A\int\!D\psi^\dagger D\psi\;\exp\int\!d^4x\;
\psi^\dagger(x)\left[i\dd+iMU^{\gamma_5}(x)\right]\psi(x).
\la{Zna}\eeq
One should remember, however, to cut the quark loop integrals
at $k\approx 1/\bar\rho\approx 600\;{\rm MeV}$. Notice that there is
no kinetic energy term for pions: it appears after one integrates
over the quark loop, see below. Summation over color is assumed in the
exponent of \Eq{Zna}.

\Eq{Zna} defines a simple local field theory though it is
still a highly non-trivial one. Its main properties will be established
in the next section. We shall see that it reproduces the known
properties of the E$\chi$L discussed above. After that we shall move
to building the nucleon from \Eq{Zna}
\footnote{We have been asked about the relation
of this low-energy theory with that suggested by Manohar and Georgi
\cite{MG}. One can redefine the quark fields

\beq
\psi\rightarrow \psi^\prime=\exp(i\pi^A\lambda^A\gamma_5/2)\psi,
\;\;\;\;\;
\psi^\dagger\rightarrow \psi^{\dagger\prime}=\psi^\dagger
\exp(i\pi^A\lambda^A\gamma_5/2),
\la{changevar}\eeq
and rewrite the lagrangian in \ur{Zna} as

\beq
{\cal L}=\psi^{\dagger\prime}(i\dd+\Dirac V + \Dirac A\gamma_5
+iM)\psi^\prime
\la{newecl}\eeq
with
\beq
V_\mu=\frac{i}{2}(\xi\partial_\mu\xi^\dagger
+\xi^\dagger\partial_\mu\xi),\;\;\;\;
A_\mu=\frac{i}{2}(\xi\partial_\mu\xi^\dagger
-\xi^\dagger\partial_\mu\xi),\;\;\;\;
\xi=\exp(i\pi^A\lambda^A/2)=U^{1/2},
\la{defs}\eeq
which resembles closely the effective lagrangian of Manohar and Georgi
(the effective chiral lagrangian in a similar form has been
independently suggested in Ref.(7)).

The crucial difference is that Manohar and Georgi have added an
explicit kinetic energy term
$F_\pi^2\Tr(\partial_\mu U^\dagger \partial_\mu U)/4$
on top of \Eq{newecl}. This seems to be unnecessary as the
kinetic energy term arises from quark loops, see the next section.}.

\section{Properties of the effective chiral lagrangian}

Properties of effective theories of quarks interacting with various
meson fields have been studied by several authors in the 80's, most
notably by Volkov and Ebert \cite{Volkov} and Dhar, Shankar and
Wadia.$\,$\cite{Dhar} The fact that integrating over quarks one gets, in
particular, the Wess--Zumino term has been first established
by Diakonov and Eides,\cite{DE} see also below.

Integrating over the quark fields in \Eq{Zna} one gets an
effective chiral lagrangian (E$\chi$L):

\beq
S_{{\rm eff}}[\pi]=-N_c\ln\det\left(i\dd+iMU^{\gamma_5}\right).
\la{Seff}\eeq
The Dirac operator in \Eq{Seff} is not hermitean:

\beq
D=i\dd+iMU^{\gamma_5},\;\;\;\;\;D^\dagger=i\dd-iMU^{\gamma_5\dagger},
\la{herm}\eeq
therefore the effective action has an imaginary part. The real part
can be defined as
\[
{\rm Re}\, S_{{\rm eff}}[\pi]=
-\frac{N_c}{2}\ln\det\left(\frac{D^\dagger D} {D_0^\dagger
D_0}\right),
\]
\beq
D^\dagger D=-\partial^2+M^2-M(\dd U^{\gamma_5}),\qquad
D_0^\dagger D_0=-\partial^2+M^2.
\la{ReS}\eeq

In the next two subsections we establish the properties of the
real and imaginary parts of the E$\chi$L separately, following
Ref.(24).

\subsection{Derivative expansion and interpolation formula}

There is no general expression for the functional \ur{Seff} for
arbitrary pion fields. For certain pion fields the functional
determinant \ur{Seff} can be estimated numerically, see section 5.
However, one can make a systematic expansion of the E$\chi$L in increasing
powers of the derivatives of the pion field, $\partial U$. It is called
long wave-length or derivative expansion. In fact, one can do it
better and expand the real part of the E$\chi$L in powers of

\beq
\frac{pM}{p^2+M^2}\:(U-1)
\la{interpopar}\eeq
where $p$ is the characteristic momentum of the pion field.
This quantity becomes small in three limiting cases: ({\em i}) small
pion fields, $\pi^A(x)\ll 1$, with arbitrary momenta, ({\em ii})
arbitrary pion fields but with small gradients or momenta, $p\ll M$,
({\em iii}) arbitrary pion fields and large momenta, $p\gg M$. We see
thus that expanding the E$\chi$L in this parameter one gets accurate
results in three corners of the Hilbert space of pion fields. For that
reason we call it {\em interpolation} formula.$\,$\cite{DPP2} Our
experience is that its numerical accuracy is quite good for more or
less arbitrary pion fields, even if one uses only the first term of
the expansion in \ur{interpopar}, see below.

The starting point for both expansions is the following formal
manipulation with the real part of the E$\chi$L \ur{ReS}.
The first move is to use the well-known formula, $\ln\det [{\rm
operator}] = \Sp\ln [{\rm operator}]$, where $\Sp$ denotes a
functional trace. One can write:
\bea
\nonumber
{\rm Re}\, S_{{\rm eff}}[\pi]\!\!&=&\!\!
-\frac{N_c}{2}\ln\det\left[1
-(-\partial^2+M^2)^{-1}M(\dd U^{\gamma_5})\right]\\
\nonumber
\!\!&=&\!\!-\frac{N_c}{2}\Sp\ln\left
[1-(-\partial^2+M^2)^{-1}M(\dd U^{\gamma_5})\right]
\eea
\bea
\nonumber
\!\!\!\!\!\!&\!\!=\!\!&\!\!-\frac{N_c}{2}
\int\!d^4x\int\!\frac{d^4k}{(2\pi)^4}\:
e^{-ik\cdot x}\Tr\ln\left[1-(-\partial^2+M^2)^{-1}M
(\dd U^{\gamma_5})\right]e^{ik\cdot x}\\
\la{splog}
\!\!\!\!\!\!&\!\!=\!\!&\!\!-\frac{N_c}{2}\!\int\!\!d^4x\!\int\!\!
\frac{d^4k}{(2\pi)^4}
\Tr\ln\!\left[1-(k^2+M^2-2ik\cdot\partial-\partial^2)^{-1}\!
M(\dd U^{\gamma_5})\right]\!\!\cdot\!\!1.
\eea
In going from the second to the third line we have written down
explicitly what does the functional trace $\Sp$ mean: take
matrix elements of the operator involved in a complete basis
(here: plane waves, $\exp (ik\cdot x)$), sum over all states (here:
integrate over $d^4k/(2\pi)^4$) and take the trace in $x$. `$\Tr$'
stands for taking not a functional but a usual matrix trace, in our
case both in flavor and Dirac bispinor indices. In going from the
third to the last line we have dragged the factor $\exp (ik\cdot x)$
through the operator, thus shifting all differential operators
$\partial\rightarrow\partial+ik$. We have put a unity at the end
of the equation to stress that the operator is acting on unity, in
particular, it does not differentiate it. Above is a standard
procedure for dealing with functional determinants.$\,$\cite{DPY}

The last line in \Eq{splog} can be now expanded in powers
of the derivatives of the pion field: it arises from expanding
\ur{splog} in powers of $\dd U^{\gamma_5}$ and of
$2ik\cdot\partial+\partial^2$. The first non-zero term has two
derivatives,

\[
{\rm Re}\, S_{{\rm eff}}^{(2)}[\pi]=\frac{N_c}{4}
\int\!d^4x\int\!\frac{d^4k}{(2\pi)^4}\:\Tr\left(\frac{M\dd U^{\gamma_5}}
{k^2+M^2}\right)^2
\]
\beq
=\frac{1}{4}\int\!d^4x\,\Tr\left(\partial_\mu U^\dagger\partial_\mu U
\right)
\cdot 4N_c \int\!\frac{d^4k}{(2\pi)^4}\:\frac{M^2}{(k^2+M^2)^2}.
\la{S2}\eeq

It is the kinetic energy term for the pion field or, better to say,
the Weinberg chiral lagrangian: atually it contains all powers of the
pion field if one substitutes $U(x)=\exp(i\pi^A(x)\lambda^A/F_\pi)$,
see \Eq{L2}. The proportionality coefficient (the last factor in
\Eq{S2}) is called $F_\pi^2$, experimentally, $F_\pi\approx 94\:MeV$.
The last factor in \Eq{S2} is logarithmically divergent; to make it
meaningful we have to recall that we have actually simplified the
theory as given by \Eq{Znash} when writing it in the local form
\ur{Zna}. Actually, the dynamical quark mass $M$ is momentum-dependent;
it cuts the logarithimically divergent integral at $k\approx
1/\bar\rho$. Using the numerical values of $\bar\rho\approx 600\:MeV$
and $M\approx 350\:MeV$ according to what follows from
instantons$\,$\cite{DP3} we find

\beq
F_\pi^2=4N_c \int\!\frac{d^4k}{(2\pi)^4}\:\frac{M^2}{(k^2+M^2)^2}
\approx \frac{N_c}{2\pi^2}M^2\ln\frac{1}{M\bar\rho}
\approx (100\:MeV)^2
\la{Fpi1}\eeq
being not a bad approximation to the experimental value of $F_\pi$.
One can at this point adjust the ultra-violet cutoff $1/\bar\rho$
to fit exactly the experimental value of $F_\pi$. Actually, the
two-derivative term is the only divergent quantity in the E$\chi$L:
higher derivative terms are all finite.

A more standard way to present the two-derivative term is by using
hermitean $N_f\times N_f$ matrices $L_\mu=iU\partial_\mu U^\dagger$.
One can rewrite \Eq{S2} as

\beq
{\rm Re}\, S_{{\rm eff}}^{(2)}[\pi]
=\frac{F_\pi^2}{4}\int\!d^4x\, \Tr L_\mu L_\mu,\qquad
L_\mu=iU\partial_\mu U^\dagger.
\la{S21}\eeq

The next, four-derivative term in the expansion of ${\rm Re}S_{{\rm eff}}$
is (note that the metric is Euclidean)

\beq
{\rm Re}\, S_{{\rm eff}}^{(4)}[\pi]=-\frac{N_c}{192\pi^2}
\int\!d^4x\,\left[2\,\Tr(\partial_\mu L_\mu)^2
+\Tr L_\mu L_\nu L_\mu L_\nu\right].
\la{S4}\eeq

These terms describe, in particular, the $d$-wave $\pi\pi$ scattering
lengths, and other observables. They can be compared with the
appropriate terms of the general Gasser--Leutwyler expansion, see
\Eq{L41}. In notations of that equation we obtain
\bea
\nonumber
L_2&=&\frac{N_c}{192\pi^2}=1.58\cdot 10^{-3},\qquad {\it vs.}
\qquad (1.35\pm 0.3)\cdot 10^{-3}\;({\rm exper.})\\
\nonumber
2L_2+L_3&=&0,\qquad\qquad\qquad
({\rm experimentally\;compatible\;with\; 0}),
\eea
being compatible with the experimental data. The
next-to-next-to-leading six derivative terms following from \Eq{splog}
have been computed in,$\,$\cite{Z,PV} however, a detailed comparison
with phenomenology is still lacking here.

We would like to mention an interesting paper \cite{PolVer} where
the derivative expansion of the E$\chi$L has been obtained from the
Lovelace--Shapiro dual resonance model. For reasons not fully
appreciated this model for the $\pi\pi$ scattering gives
coefficients in front of the 4-derivative terms numerically close to
those following from \Eq{S4}, but there is a discrepancy at the
6-derivative level.

We now turn to the interpolation formula promised in the beginning of
this subsection. One can start from the last line in \Eq{splog} and
expand it in powers of $M(\dd\Ug)$. It is clear that the actual
expansion parameter will be \ur{interpopar}. In the first non-zero
order we get \cite{DPP2}
\bea
\nonumber
{\rm Re}\, S_{{\rm eff}}^{{\rm interpol}}[\pi]
&=&\frac{N_c}{4}\!\int\!d^4x\!\!\int\!
\frac{d^4k}{(2\pi)^4}\:\Tr\!\left[\frac{1}{(k+i\partial)^2+M^2}
M(\dd U^{\gamma_5})\right.\\
\la{1interpol}
&\times &
\left.\frac{1}{(k+i\partial)^2+M^2}M(\dd U^{\gamma_5})
\right].
\eea
It will be convenient now to pass to the Fourier transform of the
$U(x)$ field understood as a matrix,

\beq
U(p)=\int\!d^4x\:e^{ip\cdot x}\left[U(x)-1\right].
\la{FTU}\eeq
The partial derivatives appearing in \Eq{1interpol} act on the exponents
of the Fourier transforms of $U,U^\dagger$ and become corresponding
momenta. As a result we get
\bea
\nonumber
{\rm Re}\, S_{{\rm eff}}^{{\rm interpol}}[\pi]
&=&\frac{1}{4}\int\!\frac{d^4p}{(2\pi)^4}
\:p^2\,\Tr\!\left[U^\dagger(p) U(p)\right]\\
\la{2interpol}
&\times& 4N_c\int\!\frac{d^4k}{(2\pi)^4}\,
\frac{M^2}{\left[(k-\frac{p}{2})^2+M^2\right]
\left[(k+\frac{p}{2})^2+M^2\right]}\:.
\eea
At $p\rightarrow 0$ the last factor becomes $F_\pi^2$ (cf. \Eq{Fpi1}),
and \Eq{2interpol} is nothing but the first term in the derivative
expansion, \Eq{S2}. However, \Eq{2interpol} also describes correctly
the functional $S_{{\rm eff}}[\pi]$ for rapidly varying pion fields
(with momenta $p\gg M$) and for small pion fields of any momenta,
when one can anyhow expand \Eq{splog} in terms of $\pi^A(x)$ and hence
in $U(x)-1$. The logarithmically divergent loop integral in
\Eq{2interpol} should be regularized, as in \Eq{Fpi1}.

Similarly, one can get the next term in the `interpolation' expansion
which will be quartic in $U(p)$, however our experience tells us that
already \Eq{2interpol} gives a good approximation to the E$\chi$L for
most pion fields.

\subsection{The Wess--Zumino term and the baryon number}

We now consider the imaginary part of $S_{{\rm eff}}[\pi]$. The first
non-zero term in the derivative expansion of ${\rm Im}\,
S_{{\rm eff}}[\pi]$ is \cite{Dhar} the Wess-Zumino
term.$\,$ \cite{WZ} It cannot be written as a $d=4$ integral
over a local expression made of the unitary $U(x)$ matrices, however
the variation of the Wess--Zumino term is local. For this reason let us
consider the variation of ${\rm Im}\, S_{{\rm eff}}[\pi]$ in respect to
the pion matrix $U(x)$. We have \cite{DPP2}
\bea
\nonumber
\delta\:{\rm Im}\, S_{{\rm eff}}[\pi]
&=& -N_c\:\delta\:{\rm Im}\ln\det D
=\frac{iN_c}{2}\Sp\left(\frac{1}{D}\delta D-\frac{1}{D^\dagger}
\delta D^\dagger\right)\\
\la{imsplog1}
&=&\frac{iN_c}{2}\Sp\left[(D^\dagger D)^{-1}D^\dagger\delta D
-(DD^\dagger )^{-1}D^\dagger \delta D^\dagger\right].
\eea
Now one can put in explicit expressions for $D,D^\dagger$ from
\Eqs{herm}{ReS}. The aim of this excercise is to get $\dd U$ in the
denominators so that an expansion in this quantity similar to that
of the previous subsection could be used.

Using the Dirac algebra relations (in Euclidean space)
\bea
\nonumber
\{\gamma_\mu,\gamma_\nu\}&=&2\delta_{\mu\nu},\qquad
\gamma_\mu^\dagger=\gamma_\mu,\qquad
\gamma_5=\gamma_1\gamma_2\gamma_3\gamma_4=\gamma_5^\dagger,\\
\la{diracalg}
\{\gamma_5,\gamma_\mu\}&=&0,\qquad
\gamma_5^2=\gamma_5,\qquad
\Tr(\gamma_5\gamma_\alpha\gamma_\beta\gamma_\gamma\gamma_\delta)
=4\epsilon_{\alpha\beta\gamma\delta},
\eea
one gets after expanding \Eq{imsplog1} in powers of $\dd \Ug$
the first non-zero term

\beq
\delta\, {\rm Im}\, S_{{\rm eff}}[\pi] = \frac{iN_c}{48\pi^2}
\int\!d^4x\:\epsilon_{\alpha\beta\gamma\delta}
\Tr\left(\partial_\alpha U^\dagger\partial_\beta U\partial_\gamma
U^\dagger \partial_\delta U\:U^\dagger\delta U\right).
\la{imsplog2}\eeq
It can be easily checked that this expression coincides with the
variation of the Wess--Zumino term written in the form a $d=5$
integral \cite{W}
\bea
\nonumber
&{\rm Im}&\!\!\!S_{{\rm eff}}[\pi]\\
\nonumber
&=&\!\!\frac{iN_c}{240\pi^2}
\int\!d^5x\:\epsilon_{\alpha\beta\gamma\delta\epsilon}\,
\Tr\Bigl(\!U^\dagger\partial_\alpha U\Bigr)
\Bigl(\!U^\dagger\partial_\beta U\Bigr)
\Bigl(\!U^\dagger\partial_\gamma U\Bigr)
\Bigl(\!U^\dagger\partial_\delta U\Bigr)
\Bigl(\!U^\dagger\partial_\epsilon U\Bigr)\\
\la{WZ}
&+&{\rm higher\;\; derivative\;\; terms}.
\eea
The integrand in \Eq{WZ} is a full derivative, however,
to write it explicitly one would need some parametrization of the
unitary matrix $U$. The expansion of \Eq{WZ} starts from the fifth
power of $\pi^A(x)$ (it is non-zero only if $N_f\ge 3$, see
subsection 3.2). It is important that, similar to
${\rm Re}\, S_{{\rm eff}}$, the imaginary part is also an infinite
series in the derivatives.

The E$\chi$L \ur{Seff} is invariant under vector flavor-singlet
transformations. Therefore, there should be a corresponding conserved
Noether baryon current, $B_\mu$. This current is associated with the
imaginary part of $S_{{\rm eff}}$ only; since ${\rm Im}\,S_{{\rm eff}}$
is an infinite series in the derivatives so is the associated Noether
current $B_\mu$. For the Wess--Zumino term \ur{WZ} the corresponding
charge is \cite{W}
\beq
B=-\frac{1}{24\pi^2}\!\int\!\!d^3\vec{x}\,\epsilon_{ijk}
\Tr\Bigl(\!U^\dagger\partial_i U\Bigr)
\Bigl(\!U^\dagger\partial_j U\Bigr)
\Bigl(\!U^\dagger\partial_k U\Bigr)
+\;{\rm higher\; derivative\; terms}.
\la{windn}\eeq
The explicitly written term is the winding number of the field
$U(x)$. Let us briefly explain this notion.

If $\pi^A(\vec{x})\rightarrow 0$ at spatial infinity so that
$U(\vec{x})\rightarrow 1$ in all directions, one can say that the
spatial infinity is just one point. \Eq{windn} gives then the winding
number for the mapping of the three-dimensional sphere $S^3$ (to which
the flat $d=3$ space is topologically equivalent when $\infty$ is one
point) to the parameter space of the $SU(N_f)$ group. In case $N_f=2$
the parameter space is also $S^3$ so that the mapping is $S^3\mapsto
S^3$. The topologically non-equivalent mappings $U(\vec{x})$, {\em
i.e.} those which can not be continuously deformed one to another, are
classified by their winding number, an integer analytically given by
\Eq{windn}. In case of $N_f> 2$ mathematicians prove that mappings are
also classified by integers given by the same \Eq{windn}.

There exists a prejudice that the baryon number carried by quarks in
the external pion field coincides with the winding number of that
field. Generally speaking it is not so because of the higher derivative
terms omitted in \Eq{windn}. Only if the pion field is spatially large
and slowly varying so that one can neglect the higher derivative
terms in \Eq{windn} one can say that the two coincide. Otherwise, for
arbitrary pion fields, the baryon number is not related to the winding
number:  the former may be zero when the latter is unity, and {\em vice
versa}.

To see what is going on here, let us calculate directly the baryon
number carried by quarks in an external time-independent pion field
$U(\vec{x})$.$\,$\cite{DPP2} The definition of the baryon charge
operator in the Minkowski space is

\beq
{\hat B} =\frac{1}{N_c}\int\!d^3\vec{x}\:\bar\psi\gamma_0\psi.
\la{BMink}\eeq
Passing to Euclidean space (which we prefer to work with since
functional integrals are more readily defined in Euclidean) one has
to make a substitution $\bar\psi\rightarrow -i\psi^\dagger,
\gamma_0\rightarrow\gamma_4$, so that

\beq
{\hat B} =-\frac{i}{N_c}\int\!d^3\vec{x}\:\psi^\dagger\gamma_4\psi.
\la{BEucl}\eeq
The baryon charge in the path integral formulation of the theory given
by \Eq{Zna} is then
\bea
\nonumber
B=\langle \hat B\rangle
&=&-\frac{i}{N_c}\int\!d^3\vec{x}\,
\langle\psi^\dagger\gamma_4\psi\rangle
=-i\int\!d^3\vec{x}\:\Tr\langle x|\gamma_4(i\dd+iM\Ug)^{-1}|x\rangle\\
\nonumber
&=&-i\int\!d^3\vec{x}\:\Tr\langle x_4,\vec{x}|
\frac{1}{i\partial_4+i\gamma_4
\gamma_k\partial_k+iM\gamma_4\Ug}|x_4,\vec{x}\rangle\\
\nonumber
&=&-i\int\!d^3\vec{x}\int_{-\infty}^{+\infty}\frac{d\omega}{2\pi}\,
\Tr\langle \vec{x}|\frac{1}{\omega+iH}|\vec{x}\rangle\\
\la{B1}
&=&\Sp\;\theta(-H)\;=\;\;{\rm\bf number\;\; of\;\; levels\;\; with\;\;
E<0}.
\eea
Here
\beq
H=\gamma_4\gamma_k\partial_k+M\gamma_4\Ug
\la{DirHam}\eeq
is the Dirac hamiltonian in the external time-independent pion field
$U(\vec{x})$ and $\theta$ is a step function.

\Eq{B1} is divergent since it sums up the baryon charge of the whole
negative-energy Dirac continuum. This divergence can be avoided by
subtracting the baryon charge of the free Dirac sea, {\it i.e.} with
the pion field switched out, $H_0=\gamma_4\gamma_k\partial_k
+M\gamma_4$:

\beq
B=-i\int\!d^3\vec{x}\int\frac{d\omega}{2\pi}\,\Tr
\langle \vec{x}|\frac{1}{\omega+iH}-\frac{1}{\omega+iH_0}|\vec{x}\rangle
=\Sp\left[\theta(-H)-\theta(-H_0)\right].
\la{B2}\eeq
In performing the integration over $\omega$ we have closed the $\omega$
integration contour in the upper semiplane. Had we closed it in the
lower semiplane we would obtain $-\Sp[\theta(H)-\theta(H_0)]$ which is
the same result since
$\Sp[\theta(H)+\theta(-H)-\theta(H_0)-\theta(-H_0)]=0$.

We have thus obtained a most natural result: the baryon charge of
quarks in the external pion field is the number of negative-energy
levels of the hamiltonian \ur{DirHam} (the number of the levels of the
free hamiltonian subtracted).

One can perform the gradient expansion for the baryon number
similarly to that of the real part of the E$\chi$L. To that end let
us write

\beq
B=-\int\!d^3\vec{x}\int_{-\infty}^{+\infty}\frac{d\omega}{2\pi}\,
\Tr\langle x|\frac{H}{\omega^2+H^2}-\frac{H_0}{\omega^2+H_0^2}
|x\rangle
\la{B3}\eeq
where
\beq
H^2=-\partial_k^2+M^2-M\gamma_4(\partial_k\Ug),\;\;\;\;\;
H_0^2=-\partial_k^2+M^2.
\la{H1}\eeq
Calculating the matrix element in the plane-wave basis one gets

\[
B=-\int\!d^3\vec{x}\int\frac{d\omega}{2\pi}\int\frac{d^3\vec{k}}
{(2\pi)^3}\:
\Tr\gamma_4\left[\frac{\vec{\gamma}\cdot(\vec{\partial}+i\vec{k})
+M\Ug}
{\omega^2-(\vec{\partial}+i\vec{k})^2+M^2
-M\vec{\gamma}\cdot(\vec{\partial}\Ug)}\right.
\]
\beq
\left.-\frac{\vec{\gamma}\cdot(\vec{\partial}+i\vec{k})}
{\omega^2-(\vec{\partial}+i\vec{k})^2+M^2}
\right]\cdot 1.
\la{B4}\eeq
For slowly varying fields $U(x)$ \Eq{B4} can be expanded in
powers of $\partial\Ug$ and $\partial$ (applied ultimately to
$\Ug$). Because of the $\Tr\gamma_5...$ the first non-zero
contribution arises from expanding the denominator in \Eq{B4}
to the third power of $\gamma\cdot (\partial\Ug)$. Integrals
over $\omega$ and $k$ are performed explicitly. After some simple
algebra one gets

\beq
B=-\frac{1}{24\pi^2}\int\!d^3\vec{x}\:\epsilon_{ijk}\Tr
(\partial_iU^\dagger\partial_jU\partial_kU^\dagger U)\;\;
+\;\;{\rm higher\;\;derivative\;\;terms}
\la{B5}\eeq
coinciding with \Eq{windn} derived from the Noether current
corresponding to the Wess--Zumino term
\ur{WZ}.\cite{DP7}$^{\!-\,}$\cite{NS}

It should be stressed that the baryon number carried by quarks in
the background pion field is equal to the topological winding
number of the field only if it is a slowly varying one.
Mathematically, the reason is that for slowly varying fields
one can neglect the higher derivative terms in \Eqs{windn}{B5}.
Physically, the reason is the following.$\,$\cite{DPP2} Imagine we
start from a pion field $U(x)$ whose winding number is one but whose
spatial size is tending to zero.  Such a field has no impact on
the spectrum of the Dirac hamiltonian \ur{DirHam}: it remains the
same as that of the free hamiltonian, namely it has the upper
($E>M$) and lower ($E<-M$) Dirac continua separated by the mass gap of
$2M$. The baryon number is zero.

We now (adiabatically) increase the spatial size of the pion field
preserving its winding number equal to unity. Since the winding
number is dimensionless this can always be done. At certain
critical spatial size the potential well for quarks formed by
the external pion field is wide enough so that a bound-state
level emerges from the upper continuum. With the increase of
the width of the potential well the bound-state level goes down towards
the lower Dirac continuum. Asymptotically, as one blows up the spatial
size of the pion field (always remaining in the winding number equal
unity sector) the bound-state level travels all the way through the
mass gap separating the two continua and joins the lower Dirac sea --
this is a theorem proven in Ref.(24). At this point one would
discover that there is an extra state close to the lower Dirac
continuum (as compared to the free, that is no-field case).
Therefore, one would say that the baryon number is now unity, --
in correspondence to \Eqs{windn}{B5}.

In a general case, however, the baryon number of the quark system
is the number of eigenstates of the Dirac hamiltonian \ur{DirHam}
one bothers to fill in. The role of the winding number of the
background pion field is only to guarantee that, if the spatial
size of the field is large enough, the additional bound-state
level emerging from the upper continuum is a deep one:
asymptotically it goes all the way to the lower continuum.

\section{The nucleon}

All constituent quark models start from assuming that the nearly
massless light quarks of the QCD lagrangian obtain a non-zero
dynamical quark mass $M\approx 350\;{\rm MeV}$. This is due
to the spontaneous chiral symmetry breaking (SCSB), its microscopic
driving force being, to our belief, instantons, see subsection 3.5.
Even if one does not believe in instantons as the microscopic mechanism
of SCSB one has anyhow to admit that once light quarks get a dynamical
or constituent mass $M(k)$ such quarks inevitably have to interact
with Goldstone pions, as explained in subsection 3.4.

How strong, numerically, is this interaction?
Expanding $\Ug$ \ur{Ug} to the first
order of the pion field we get from \Eq{chirla} that the linear coupling
of pions to constituent quarks is, numerically, quite large:
\bea
\nonumber
{\cal L}^{{\rm int}}&=&ig_{\pi qq}\,\pi^A\,
\left(\bar\psi\gamma_5\lambda^A\psi\right),\\
\la{coupling}
g_{\pi qq}&=&\frac{M}{F_\pi}\simeq 4.
\eea
We would like to emphasize that this is a model-independent
consequence of saying that quarks get a constituent mass.

At distances between quarks of the order of 0.5 fm
typical for interquark separations inside nucleons, neither the
one-gluon exchange nor the supposed linear potential are as large
as the chiral forces. Therefore, it is worthwhile to investigate
whether the chiral forces alone are able to bind the constituent
quarks inside nucleons.

\subsection{A note on confinement}

The generally accepted confinement scenario is, verbally, very simple.
In the pure glue world (without light quarks) it is expected that the
Coulomb potential at small separations between static probe quarks is
replaced at large separations by a linear potential growing {\it ad
infinitum}.  A constant 14 ton force is supposed to be acting between
probe quarks even if one is on the Earth and the other one is on the
Moon. This is a very radical departure from forces decaying with
distance, which have been known to physics so far. If proven correct,
it would be one of the most remarkable discoveries in physics of all
times.

There have been several proposals of microscopic scenarios which could
lead to the linear potential, such as the condensation of magnetic
monopoles or a vacuum filled by random $Z(N_c)$ vortices. We are not
in a position to discuss them here but would like to note that,
despite enormous efforts in the last 25 years, none of them have been
worked out mathematically so far, at least at the semi-quantitative
level the instanton vacuum has been constructed. The only theoretical
model we know of where one gets a linear potential in more than two
dimensions is the old example by A.Polyakov$\,$\cite{Pol} of the $d=2+1$
Georgi--Glashow model, but there the non-Abelian color group is broken
down to $U(1)$ by the Higgs condensate. Therefore, our belief
that the asymptotic linear potential takes place in pure gluodynamics
is grounded not on theory but mainly on the results of lattice
simulations.

When one looks carefully into how lattice practitioners come to the
conclusion about the linear potential one finds several disturbing
details. Some of them are discussed in our recent paper$\,$\cite{DP9}.
We mention just one here. The standard way to extract the quark
potential on a lattice is by measuring large Wilson loops $W(r,t)$.
One side of the loop is called ``$r$'' (separation), the other one is
``$t$'' (time). If one wants to measure the potential $V(r)$
at separation $r$ one needs to take $t\gg r$, otherwise the
measurement will be contaminated by many excited states. However, such
measurements are practically impossible in the interesting range of
$r\ge 1\;{\rm fm}$ because the average Wilson loop becomes so small
there that measuring it would require statistics many orders of
magnitude exceeding that available with today computers.$\,$\cite{DP9}

Therefore, this route is abandoned. Instead, Wilson loops are measured
in the opposite limit, $t\ll r$. The hope is that the slope
$-\partial\ln W(r,t)/\partial t$ reaches its asymptotic ($t\to\infty$)
value $V(r)$ already at small values of $t$. One tries to assist
this early convergence by taking `smeared' links along the $r$ sides
of the loop, which are expected to have better overlap with the ground
state of the potential. One of the best (if still not the best)
measurement of that kind \cite{Wup} is presented in Fig. 1.

\begin{figure}
\centerline{\psfig{figure=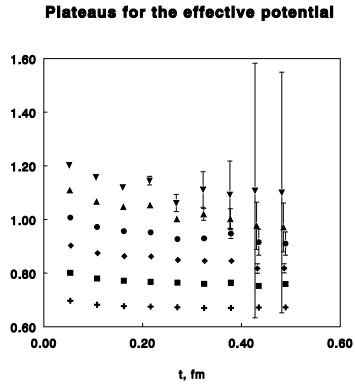,height=5.0cm}}
\caption{Effective potential
$V_{{\rm eff}}(r,t)=\ln\left[W(r,t)/W(r,t+a)\right]$ as function of $t$
at different values of $r$ from Ref.(35).
The data points correspond (from bottom to top) to
$r=0.65,\;0.98,\;1.30,\;1.62,\;1.95\;{\rm and}\; 2.25\;{\rm fm}$.
The inverse coupling used $\beta=2.635$ corresponds to the lattice
spacing $a=0.0541\;{\rm fm}$. Courtesy G.Bali.}
\label{fig:picture0}
\end{figure}

One can see from Fig. 1 that the quality of the plateaus at
$r\leq 1\;{\rm fm}$ is quite good though the data tend to slope down
as $t$ increases. At larger $r$ this trend becomes more pronounced,
until the error bars explode so that one can hardly get to any
conclusions at all. The procedure of extracting the potential is
detailed in Ref.(35) but basically it follows from the data in
Fig. 1. Thus, the extraction of the linear potential at large $r$
fully relies on the faith in that the `plateaus' of Fig. 1 are not
going to slope down as $t$ increases from the region $t\ll r$ where
they are measured to $t\gg r$ where they should be measured.

As explained in \cite{DP9} this reliance is potentially dangerous.
At $t\ll r$ one can view the Wilson loop's long side $r$ as `time' and
the short side $t$ as `separation'; then the linear dependence of
$\ln W(r,t)$ on the long side $r$, i.e. the `linear potential'
is built in, whereas it is exactly what is so demanding to prove.
This remark becomes even more disturbing in view of the fact that
the linear potential is also observed in two cases where the potential
{\em has} to flatten owing to screening: (i) adjoint sources and (ii)
fundamental sources but with light dynamical fermions. In all cases
the same procedure of extracting the potentials has been used, namely
from measuring Wilson loops at $t\ll r$, and no clear signals of
flattening has been observed.

In our mind, the problem of demonstrating the asymptotic linear
potential both theoretically and `experimentally' (on the lattice)
remains as challenging as it was 25 years ago.

Whatever the outcome of those very important studies, the world we live
in is not pure glue: we have light quarks. Light quark-antiquark pairs
produced from the vacuum eventually screen any rising potential. In the
chiral limit pions are massless, hence the screening starts at the
point where the potential exceeds zero energy. In reality the potential
cannot exceed $2m_\pi=280\;{\rm MeV}$ which is reached, in the standard
(Coulomb + linear potential) parametrization at a rather small
$0.3\;{\rm fm}$ separation. Nucleon is a stationary state; there is an
infinite time for any string exceeding that length to decay into pions.
Therefore, it seems to be senseless to employ large confining
potentials in a realistic model of nucleons.

Why, then, do not quarks `get away' but exist only in bound states?
A simple explanation can be provided by the SCSB itself. As a result of
SCSB originally massless or nearly massless quarks get a dynamical mass
$M\approx 350\;{\rm MeV}$, and pions become the lightest
excitations in the spectrum. Therefore, light quarks need not exist as
asymptotic states: instead of producing a quark-antiquark pair it
is energetically favorable to produce one or several pions.
Mathematically, it would correspond to the quark propagator with
momentum-dependent mass, having singularities on the second Riemann
sheet under the cut starting from the pion threshold.

What about confinement of heavy quarks ($c,b,t$)? Let us imagine,
for the sake of an argument, that the static potential between probe
quark sources in a pure glue theory is not rising to infinity but
levels off at some $V_\infty=2\Delta M_\infty$. If $\Delta M_\infty$
happens to be larger than approximately the light constituent quark
mass $M$ \footnote{This condition is readily met e.g. in the instanton
vacuum.\cite{DP9}} heavy quarks would be unstable under decay to
$D$ or $B$ mesons. Heavy quarks can be thus confined too. The
case is to some extent similar to electrodynamics with charges $Z>137$:
such particles are unstable under a spontaneous production of $e^+e^-$
pairs and therefore cannot exist as asymptotic states.$\,$\footnote{The
``$Z>137$'' scenario of confinement has been advocated by
V.N.Gribov.\cite{Gr}}

\subsection{A qualitative picture of the nucleon}

Leaving aside those intriguing problems, we concentrate on the lowest
state with baryon number one, {\em i.e.} the nucleon. As mentioned
above the interquark separations in the ground-state nucleon are
moderate (order of $0.5\;{\rm fm}$) and it is worthwhile asking whether
the simple E$\chi$L \ur{Zna} is capable of explaining the basic
properties of the ground-state nucleon. Notice that the expected
typical momenta of quarks inside the nucleon are of the order of
$M\approx 350\;{\rm MeV}$, that is in the domain of applicability of
the low-momentum effective theory \ur{Zna}, see subsection 3.6.

The chiral interactions of constituent quarks in the 3-quark nucleon,
as induced by the effective theory \ur{Zna}, are schematically shown
in Fig. 2, where quarks are denoted by solid and pions by dashed lines.
Notice that, since there is no tree-level kinetic energy for pions in
\Eq{Zna}, the pion propagates only through quark loops. Quark
loops induce also many-quark interactions indicated in Fig.2 as well.
We see that the emerging picture is, unfortunately, rather far from
a simple one-pion exchange between the constituent quarks: the
non-linear effects in the pion field are not at all suppressed.

\begin{figure}
\centerline{\psfig{figure=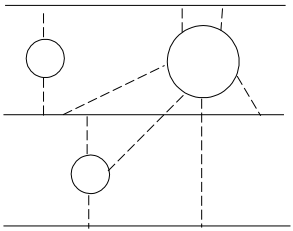,height=3.0cm}}
\caption{Quarks in a nucleon interacting via pion fields.}
\label{fig:picture2}
\end{figure}

At this point one may wonder: isn't the resulting theory as complicated
as the original QCD itself? The answer is no, the effective low-energy
theory is a simplification as compared to the original
quark-gluon theory, because it deals with adequate degrees of freedom.
Let us imagine that one would like to describe `low-energy' properties
of solid states, for example superconductivity. Would working with the
underlying theory (QED) be helpful? Not too much. We know that the
microscopic theory leads, under certain conditions, to the
rearrangement of atoms into a lattice, so that translational symmetry
is spontaneously broken. As a result the Goldstone bosons appear
(here: phonons), and electrons get a dynamical mass different from
the input one. The most important forces are due to phonon exchange
between electrons: in fact they are driving superconductivity in the
BCS theory. Playing with this analogy, nucleon is like a polaron (a
bound state of electrons in the phonon field), rather than a
positronium state in the vacuum. After chiral symmetry is broken
we deal with a `metal' phase rather than with the vacuum one, and one
has to use adequate degrees of freedom to face this new situation.

Instantons play the role of the bridge between the
microscopic theory (QCD) and the low-energy theory where one neglects
all degrees of freedom except the Goldstone bosons and fermions with
the dynamically-generated mass. Instantons do the most difficult
part of the job: they explain why atoms in metals are arranged into a
lattice, what is the effective mass of the electron and what is the
strength of the electron-phonon interactions. However, one can
take an agnostic stand and say: I don't care how 350 MeV is obtained
from the microscopic $\Lambda_{{\rm QCD}}$ and why do
atoms form a lattice in the metals: I just know it happens.
To such a person we would advise to take the low-energy theory
\ur{Zna} at face value and proceed to the nucleon.

A considerable technical simplification is achieved in the limit of
large $N_c$. For $N_c$ colors the number of constituent quarks in a
baryon is $N_c$ and all quark loop contributions are also proportional
to $N_c$, see section 4. Therefore, at large $N_c$ one can speak about
a {\em classical self-consistent} pion field inside the nucleon:
quantum fluctuations about the classical self-consistent field are
suppressed as $1/N_c$. The problem of summing up all diagrams of the
type shown in Fig. 2 is thus reduced to finding a classical pion field
pulling $N_c$ massive quarks together to form a bound state.

\subsection{Nucleon mass: a functional of the pion field}

Let us imagine that there is a classical time-independent pion
field which is strong and spatially wide enough to make a  bound-state
level of the Dirac hamiltonian \ur{DirHam} for massive quarks, call its
energy $E_{{\rm level}}$. We fill in this level by $N_c$ quarks in the
antisymmetric state in color, thus obtaining a baryon number one
state, as compared to the vacuum. The interactions with the background
chiral field are color-blind, so one can put $N_c$ quarks
on the same level; the fact that one has to put them in an
antisymmetric state in color, {\em i.e.} in a color-singlet state,
follows from Fermi statistics.

One has to pay for the creation of this trial pion field, however.
Call this energy $E_{{\rm field}}$. Since there are no direct terms
depending on the pion field in the low-momentum theory \ur{Zna}
the only origin of $E_{{\rm field}}$ is the fermion determinant
\ur{Seff} which should be calculated for time-independent field
$U(\vec{x})$. It can be worked out with a slight modification of
the equations of section 4. We have:$\,$\cite{DPP2}
\bea
\nonumber
S_{{\rm eff}}[\pi]&=&-N_c\:\ln\;\det\left(\frac{D}{D_0}\right)\\
\nonumber
&=&-N_c\;\Sp\left[\ln(i\partial_t+iH)-\ln(i\partial_t+iH_0)\right]\\
\la{Efi1}
&=&-TN_c\int\!\frac{d\omega}{2\pi}\,\Sp\left[\ln(\omega+iH)
-\ln(\omega+iH_0)\right],
\eea
where $H$ is the Dirac hamiltonian \ur{DirHam} in the stationary pion
field, $H_0$ is the free hamiltonian and $T$ is the (infinite) time
of observation. Using an important relation \footnote{It follows from
taking the matrix trace of the hamiltonian \ur{DirHam}. It should be
kept in mind, though, that such a naive derivation can be potentially
dangerous because of anomalies in infinite sums over levels. However,
it can be checked that in this particular case there are no anomalies
and the naive derivation is correct.}

\beq
\Sp(H-H_0)=0
\la{Spid}\eeq
(telling us that the sum of all energies, with their signs, of
the Dirac hamiltonian \ur{DirHam} is the same as for the free case)
one can integrate in \Eq{Efi1} by parts and get
\bea
\nonumber
S_{{\rm eff}}[\pi]&\equiv& TE_{{\rm field}}
=TN_c\int\!\frac{d\omega}{2\pi}\,\Sp\left[
\frac{\omega}{\omega+iH}-\frac{\omega}{\omega+iH_0}\right]\\
\la{Efi2}
&=&TN_c\sum_{E_n^{(0)}<0}\left(E_n-E_n^{(0)}\right).
\eea
Going from the first to the second line we have closed the $\omega$
integration contour in the upper semiplane; owing to the trace relation
\ur{Spid} closing it in the lower semiplane would produce the same
result.

We see that the price $E_{{\rm field}}$ one pays for a creation
of the time-independent pion field coincides with the aggregate energy
of the lower Dirac continuum in that field. The energy of the additional
level emerging from the upper continuum, which one has to fill in to
get the baryon number one state, $E_{{\rm level}}$, should be added to
get the total nucleon mass. This simple scheme \cite{DP8,DPP2}
is depicted in Fig. 3. Naturally, the mass of the nucleon should be
counted from the vacuum state corresponding to the filled levels of
the free lower Dirac continuum. Therefore, the (divergent) aggregate
energy of the free continuum should be subtracted, as in \Eq{Efi2}.

We have thus for the nucleon mass:

\beq{\cal M}_N=\mathrel{\mathop{{\rm min}}\limits_{\{\pi^A(\vec{x})\}}}
\left(N_cE_{{\rm level}}[\pi]+E_{{\rm field}}[\pi]\right).
\la{nuclmass}\eeq

Both quantities, $E_{{\rm level}}$ and $E_{{\rm field}}$, are functionals
of the trial pion field $\pi^A(\vec{x})$. The classical self-consistent
pion field is obtained from minimizing the nucleon mass \ur{nuclmass}
in $\pi^A(\vec{x})$. It is called the {\it soliton} of the
non-linear functional \ur{nuclmass}, hence the {\bf Chiral
Quark-Soliton Model}. An accurate derivation of \Eq{nuclmass} from the
path-integral representation for the correlation function of Ioffe
currents is presented in Ref.(24). It solves the problem of
summing up all diagrams of the type shown in Fig. 2 in the large-$N_c$
limit.

The idea that a sigma model with pions coupled directly
to constituent quarks, can be used to build the nucleon soliton has
been first suggested by Kahana, Ripka and Soni \cite{KRS} and
independently by Birse and Banerjee.$\,$\cite{BB} We would call them the
authors of the Chiral Quark-Soliton Model. Technically, however, in
these references an additional {\em ad hoc} kinetic energy term for
pion fields has been used, leading to a vacuum instability
paradox.$\,$\cite{Soni,RK} The present formulation of the model has
been given in Ref.(37), together with the discussion of its domain of
applicability and its physical contents. A detailed theory based on the
path integral approach paving the way for calculating nucleon
observables has been presented in Ref.(24).

\begin{figure}
\centerline{\psfig{figure=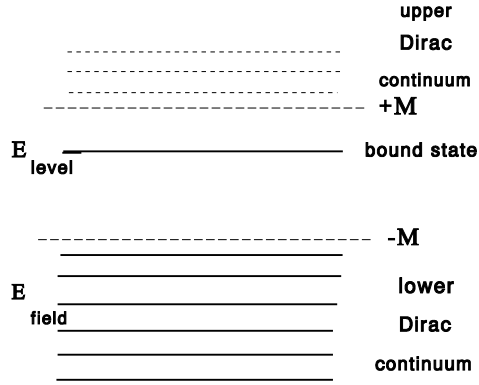,height=5.0cm}}
\caption{Spectrum of the Dirac hamiltonian in trial
pion field. The solid lines show occupied levels.}
\end{figure}

\subsection{Nucleon profile}

To find the classical pion field minimizing the nucleon mass
\ur{nuclmass} one has first of all decide on the symmetry of the
pion field. Had the field been a singlet one would take a
spherically-symmetric ansatz. However, the pion field has flavor indices
$A=1,...,N_f^2-1$. At $N_f=2$ the three components of the pion field
can be married with the three space axes. This is called the
hedgehog ansatz; it is the minimal generalization of spherical symmetry
to incorporate the $\pi^\pm=(\pi^1\pm i\pi^2)/\surd{2}$ and
$\pi^0=\pi^3$ fields:

\beq
\pi^A(\vec{x})/F_\pi=n^AP(r),\;\;\;\;\;n^A=\frac{x^A}{r},\;\;\;\;\;
r=|\vec{x}|,
\la{hedgehog}\eeq
where $P(r)$ is called the soliton profile.

The choice of the ansatz is not at all innocent: baryons corresponding
to different choices would have qualitatively different properties,
see the next subsection. The maximally-symmetric ansatz \ur{hedgehog}
will have definite consequences in applications.

In the $N_f=3$ case there are 8 components of the `pion' field, and
there are several possibilities to marry them to the space axes. The
commonly used is the `left upper corner' ansatz:$\,$\cite{W2}

\beq
U(\vec{x})\equiv \exp\left(i \pi^A(\vec{x})\lambda^A\right)
= \left(\begin{array}{cc}
\exp \left[i(\mbox{\boldmath{$n$}}
\cdot\mbox{\boldmath{$\tau$}}) P(r)\right] &
\begin{array}{c} 0\\ 0\end{array}\\ \begin{array}{cc}0\;\;&\;\;\;0
\end{array}& 1\end{array}\right),\;\;\;\;
\vec{n}=\frac{\vec{x}}{r}.
\la{leftupper}\eeq
As we shall see in the next subsection, the quantization of
rotations for this ansatz leads to the correct spectrum of the lowest
baryons.

Another $SU(3)$ ansatz \cite{Bal} discussed in the literature is
($f,g,h=1,2,3$ are the flavor indices):

\beq
U_{fg}=e^{iP_2/3}\left[\cos P_1\:\delta_{fg}+\left(e^{-iP_2}
-\cos P_1\right)n_fn_g+\sin P_1\:\epsilon_{fgh}n_h\right],
\la{bal}\eeq
where $P_{1,2}(r)$ are spherically-symmetric profile functions.
This ansatz is used to describe strangeness $-2$
dibaryons.$\,$\cite{DPPP} We shall not consider it here but
concentrate on the usual baryons for which the hedgehog ansatz
\ur{hedgehog} or \ur{leftupper} is appropriate.

Let us first discuss restrictions on the best profile function $P(r)$
which should mimimize the nucleon mass \ur{nuclmass}. What is the
asymptotics of $P(r)$ at large $r$? To answer this question one has
to know the behavior of $E_{{\rm field}}[\pi]$ for slowly varying pion
fields. Using \Eq{Efi2} as a starting point one can work out the
derivative expansion of the functional $E_{{\rm field}}[\pi]$
similar to that for the full E$\chi$L, see subsection 3.1. We have
\cite{DP8,DPP2}

\[
E_{{\rm field}}[\pi]=\frac{F_\pi^2}{4}\int\!d^3\vec{x}\,\Tr L_iL_i
-\frac{N_c}{192\pi^2}\int\!d^3\vec{x}\:\left[2\,\Tr(\partial_iL_i)^2
+\Tr L_iL_jL_iL_j\right]
\]
\beq
+\;\;{\rm higher\;\; derivative\;\; terms},
\;\;\;\;\;L_i=iU\partial_iU^\dagger.
\la{Efi3}\eeq

Substituting here the hedgehog ansatz one gets a functional of the
profile function $P(r)$; varying it one finds the Euler--Lagrange
equation valid for slowly varying profiles, in particular, for
the tail of $P(r)$ at large $r$. It follows from this equation
that $P(r)=A/r^2$ at large $r$. The second contribution to the nucleon
mass, $E_{{\rm level}}[\pi]$, does not alter this derivation since
the bound-state wave function has an exponential but not power behavior
at large $r$. Actually we get the pion tail inside the nucleon, and the
constant $A$ is related to the nucleon axial constant. This relation
is identical to the one found in the Skyrme model:$\,$\cite{ANW}

\beq
g_A=\frac{8\pi}{3}AF_\pi^2.
\la{gA}\eeq
The exponentially decreasing wave function of the bound-state level
does not change this equation, as well as the Goldberger--Treiman
relation for the pion-nucleon coupling constant,

\beq
g_{\pi NN}=\frac{g_A{\cal M}_N}{F_\pi}
=\frac{8\pi A}{3}\frac{{\cal M}_N}{F_\pi}.
\la{GT}\eeq
Furthermore, it follows from the next four-derivative term in \Eq{Efi3}
that the $1/r^4$ correction to $P(r)$ at large $r$ is
absent!$\;$\cite{DP8,DPP2} It means probably that the pion tail
inside the nucleon is unperturbed to rather short distances.

The second important question is what should we choose for $P(0)$?
As explained in subsection 3.2, a quantity which guarantees a
deeply-bound state in the background pion field is the winding number
of the field, \Eq{windn}. Substituting the hedgehog ansatz into
\Eq{windn} one gets

\beq
N_{{\rm wind}}
=-\frac{2}{\pi}\int_0^\infty\!dr\:\sin^2P(r)\frac{dP(r)}{dr}
=-\frac{1}{\pi}\left[P(r)-\frac{\sin 2P(r)}{2}\right]_0^\infty.
\la{windhedge}\eeq
Since $P(\infty)=0$ the way to make this quantity unity is to choose
$P(0)=\pi=3.141592\ldots$

An example of a one-parameter variational function satisfying the above
requirements is \cite{DP8,DPP2}

\beq
P(r)=2\:{\rm arc\,tan}\left(\frac{r_0}{r}\right)^2,\qquad
A=2r_0^2.
\la{arctg}\eeq

The Dirac hamiltonian \ur{DirHam} in the hedgehog pion field
\ur{hedgehog} commutes neither with the isospin operator $T$ nor
with the total angular momentum $J=L+S$ but only with their sum
$K=T+J$ called the `grand spin'. The eigenvalue Dirac equations for
given value of $K^2, K_3$ have been derived in Ref.(24).
Generally speaking, there appears a bound-state level with the
$K^P=0^+$ quantum numbers whose energy can be found from solving the
Dirac equations for two spherically-symmetric functions $j,h$,
\bea
\nonumber
\frac{dh}{dr}&=&-M\sin P\:h+(E_{{\rm level}}+M\cos P)\:j,\\
\la{K0}
\frac{dj}{dr}+\frac{2}{r}\:j&=&M\sin P\:j+(-E_{{\rm level}}
+M\cos P)\:h,
\eea
with the boundary conditions $h(0)=1,\; j(0)=Cr,\;
h(\infty)=j(\infty)=0$.

These equations determine one of the two contributions to the nucleon
mass, $E_{{\rm level}}$. The second contribution, namely that of
the aggregate energy of the lower Dirac continuum in the trial
pion field, which we have called $E_{{\rm field}}$, can be found
in several different ways. One way is to find the phase shifts in the
lower continuum, arising from solving the Dirac equation for definite
grand spin $K$.\cite{DPP2,DPP3} Another method is to diagonalize
the Dirac hamiltonian \ur{DirHam} in the so-called Kahana--Ripka
basis \cite{RK} written for a finite-volume spherical box. Both
methods are, numerically, rather involved. There exists a third
(approximate) method \cite{DP8,DPP2} allowing one to make an estimate of
$E_{{\rm field}}$ in a few minutes on a PC. It is based on the
interpolation formula for the E$\chi$L, see \Eq{2interpol}.

Let us discuss the qualitative behavior of $E_{{\rm level}}$ and
$E_{{\rm field}}$ with the soliton scale parameter $r_0$ assuming
for definiteness that the profile is given by \Eq{arctg}.

The trial pion field plays the role of the (relativistic) potential
well for massive quarks. The `depth' of this potential well is fixed
by the condition $P(0)=\pi$ and  is always finite: this is
related to the fact that the pion field has the meaning of angles.
The spatial size of the trial pion field $r_0$ plays the role of the
`width' of the potential well. It is well known that in three
dimensions the condition for the appearance of a bound state is
$MVr_0^2>{\rm const}$ where $V$ is the depth of the well and `const' is
a numerical constant of the order of unity depending on its concrete
shape. In our case $V\approx M$, so the condition
that the bound state appears is $r_0M=O(1)$. Therefore, at small
sizes $r_0$ there is no bound state for the Dirac hamiltonian
\ur{DirHam}, so that $E_{{\rm level}}$ coincides with the border of the
upper Dirac continuum, $E_{{\rm level}}=+M$. At certain critical value
of $r_0$ a weakly bound state emerges from the upper continuum. [For
the concrete ansatz \ur{arctg} the threshold value is $r_0M\simeq
0.5$.] As one increases $r_0$ the bound state goes deeper and $E_{{\rm
level}}$ monotonously decreases. At very large spatial sizes,
$r_0\rightarrow\infty$, $E_{{\rm level}}$ approaches the lower
continuum, its difference from $-M$ falling as $1/r_0^2$.$\,$\cite{DPP2}
The behavior of $E_{{\rm level}}$ as function of $r_0$ is
plotted in Fig. 4.

The monotonous decrease of $E_{{\rm level}}$ with the increase of $r_0$
is a prerogative of the trial pion field with winding number 1.
Had it been zero, $E_{{\rm level}}$ would first go down and then
start to go up, asymptotically joining back the {\em upper} continuum.
In the case of $N_{{\rm wind}}=-1$ the bound state would travel in the
opposite direction: from the lower towards the upper continuum. At
$N_{{\rm wind}}=n$ as much as $n$ levels would emerge, one by one, from
the upper continuum and travel all the way through the mass gap towards
the lower one. For the trial pion field of the hedgehog form all these
things happen exclusively for states with grand spin
$K=0$.$\,$\cite{DPP2}

Turning now to $E_{{\rm field}}$ we first notice that for large
spatial sizes of the trial pion field one can use the first term
in the derivative expansion for $E_{{\rm field}}$, see \Eq{Efi3}.
On dimension grounds one immediately concludes that
$E_{{\rm field}}\sim F_\pi^2r_0$ for large $r_0$, {\em i.e.} is
infinitely linearly rising in $r_0$. At small $r_0$ a slightly more
complicated analysis \cite{DPP2} shows that  $E_{{\rm field}}\sim
r_0^3$. On the whole, $E_{{\rm field}}$ is a monotonously rising
function of $r_0$ shown in Fig. 4.

\begin{figure}
\centerline{\psfig{figure=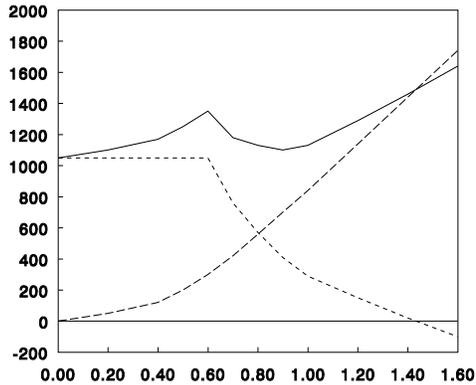,height=5.0cm}}
\caption{Nucleon mass and its constituents
as function of the soliton size. The short-dash line shows
$3E_{{\rm level}}$, the long-dash line is $E_{{\rm field}}$, the
solid line is their sum, ${\cal M}_N$.}
\end{figure}

The nucleon mass, ${\cal M}_N=3E_{{\rm level}}+E_{{\rm field}}$
(for $N_c=3$) is also plotted in Fig. 4 taken from
Refs.(37,24). One observes a non-trivial minimum for
${\cal M}_N$ corresponding to $r_0\simeq 0.98/M\simeq 0.57\:{\rm fm}$.
This is, phenomenologically, a very reasonable value, since from
\Eqs{gA}{GT} one immediately gets
$g_A\simeq 1.15\;{\rm versus}\;1.25\;{\rm (exp.)}$ and
$g_{\pi NN}\simeq 13.6\;{\rm versus}\;13.5\;{\rm (exp.)}$.
The nucleon mass appears to be ${\cal M}_N\simeq 1100\:{\rm MeV}$
with $3\,E_{{\rm level}}\simeq 370\:{\rm MeV}$,
$E_{{\rm field}}\simeq 730\: {\rm MeV}$.
Note that the `valence' quarks (sitting on the bound-state level)
come out to be very strongly bound:  their wave function falls
off as $\exp(-r/0.6\:{\rm fm})$, and about $2/3$ of the quark mass
$M\simeq 350\:{\rm MeV}$ is eaten up by interactions with the classical
pion field. Relativistic effects are thus essential.

Though the nucleon bound state appears to be somewhat higher
than the free-quark threshold, $3M\simeq 1050\:{\rm MeV}$,
there are several known corrections to it which are negative.
The largest correction to the nucleon mass is due to
taking into account explicitly the one-gluon exchange between both
`valence' and `sea' quarks; this correction is $O(N_c)$ as is the
nucleon mass itself. Numerically, it turns out to be about
$-200\:{\rm MeV}$ \cite{DJP} and seems to move the nucleon mass just into
the right place.$\,$\footnote{There exist also numerous {\em quantum}
corrections to the nucleon mass of different origin, which are of
the order of $O(N_c^0)$.  Unfortunately, it is difficult today to
treat them in a systematic fashion.}

We see thus that the `valence' quarks in the nucleon get bound
by a self-consistent pion field whose energy is given by
the aggregate energy of the negative Dirac continuum distorted
by the presence of the external field. This picture of the nucleon
interpolates between the old non-relativistic quark models (which
would correspond to a shallow bound-state level and an undistorted
negative continuum) and the Skyrme model (which would correspond to a
spatially very large pion soliton so that the bound-state level would
get close to the lower continuum and the field energy $E_{{\rm field}}$
would be given just by a couple of terms in its derivative expansion).
The reality is somewhere in between: the bound-state level is a deep
one but not as deep as to say that all the physics is in the lower
Dirac continuum.

Idelogically, this picture of the nucleon at large $N_c$ is
somewhat similar to the Thomas--Fermi picture of the atom at large
$Z$. In that case quantum fluctuations of the self-consistent
electrostatic field binding the electrons are suppressed by large
$Z$, however corrections go as powers of $Z^{-2/3}$, as contrasted to
the $N_c^{-1}$ corrections of the Chiral Quark-Soliton Model.
Therefore, the latter model is in a slightly better position in
respect to quantum corrections than the former.

Apart from using the large $N_c$ approximation (which is in fact
just a technical device needed to justify the use of the classical
pion field) the Chiral Quark-Soliton Model makes use of the
small algebraic parameter $(M\bar\rho)^2\approx 1/3$
where $1/\bar\rho$ is the UV cutoff of the E$\chi$L \ur{Zna}. [In the
instanton derivation $\bar\rho$ is the average size of instantons in
the vacuum.] This $\bar\rho$ is, roughly,
the size of the constituent quark, while the size of the nucleon is,
parametrically, $1/M$. The fact that the constituent quark picture
works so well in the whole hadron physics finds its explanation in
this small numerical parameter. [In the instanton picture it is due to
the relative diluteness of the instanton vacuum, which in its turn is
related to the `accidentally' large number (11/3) in the asymptotic
freedom law.$\,$\cite{DP1,DPW}] The small parameter
$(M\bar\rho)^2\ll 1$ makes it possible to use only quarks with
dynamically generated mass and chiral fields as the only essential
degrees of freedom in the range of momenta $k\sim M\ll 1/\bar\rho$, and
that is exactly the range of interest in the nucleon binding problem.

The above numerics have been obtained from the interpolation
formula for $E_{{\rm field}}$.$\,$\cite{DP8,DPP2} Exact calculations
of $E_{{\rm field}}$ performed in \cite{DPP3} as well as taking
more involved profiles with three variational parameters did not
lead to any significant changes in the numerics.

Following refs. \cite{DP8,DPP2} there had been many
calculations of the nucleon mass and of the `best' profile using
various regularization schemes and parameters of the chiral model,
see \cite{Review} for a review. The effective low-momenta theory
\ur{Zna} comes along with an intrinsic ultraviolet cutoff,
in the form of a momentum dependence of the
constituent quark mass, $M(k)$. In the instanton approach, it can be
shown on general grounds that this is a rapidly decreasing function at
momenta of the order of the inverse average instanton size, $1/\bar\rho
\approx 600\, {\rm MeV}$.  However, the present `state of the art' does
not allow one to determine this function accurately at all values of
momenta -- to do so, one would need a very detailed understanding of
the QCD vacuum.  This places certain restrictions on the kinds of
quantities which can sensibly be computed using the effective theory.
Those are either finite ones, which do not require an UV cutoff at all,
or quantities at most logarithmically divergent. Both type of
quantities are dominated by momenta much smaller than the UV
cutoff, $k \ll 1/\bar\rho$, so one can compute them mimicking the
fall--off of $M(k)$ by an external UV cutoff
$\Lambda\simeq 1/\bar\rho$, using  some regularization
scheme.$\,$\footnote{Recently the self-consistent field binding
the nucleon has been found using directly the non-local coupling
\ur{Znash}.$\,$\cite{BR}}
Fortunately, almost all nucleon observables belong to these two
classes. The uncertainty related to the details of the ultra-violet
regularization leads to a 15-20\% numerical uncertainty of the
results, and that is the expected accuracy of the model today.

\section{Quantum numbers of baryons}

The picture of the nucleon outlined in the previous section is
``classical'': the quantum fluctuations of the self-consistent
pion field binding $N_c$ quarks are totally ignored.
Among all possible quantum corrections to the nucleon mass
a special role belongs to the zero modes. Fluctuations of the pion
field in the direction of the zero modes cannot be considered small,
and one has to treat them exactly. Zero modes are always related to
continuous symmetries of the problem at hand. In our case there are
3 zero translational modes and a certain number of zero rotational
modes.  The latter determine the quantum numbers of baryons;
it is here that the hedgehog (or whatever) symmetry of the ansatz
taken for the self-consistent pion field becomes crucial.

A general statement is that if the chiral field $U_{{\rm cl}}(\vec{x})$
minimizes the nucleon mass functional \ur{nuclmass}, a field
corresponding to rotated spatial axes, $x_i\rightarrow O_{ij}
x_j$, or to a unitary-rotated matrix in flavor space,
$U_{{\rm cl}}\rightarrow RU_{{\rm cl}}R^\dagger$, has obviously
the same classical mass. This is because the functional \ur{nuclmass}
to be minimized is isotropic both in flavor and ordinary spaces.

Specifically for the hedgehog ansatz [see \Eq{hedgehog} for the flavor
$SU(2)$ and \Eq{leftupper} for the $SU(3)$] any spatial rotation is
equivalent to a flavor rotation. We shall first show it for a more
general case of $SU(3)$.

\subsection{Case of three flavors}

The space-rotating $3\times 3$ matrix $O_{ij}$ can be written as

\beq
O_{ij}=\frac{1}{2}\Tr (S\tau_i S^\dagger\tau_j)
\la{Oij}\eeq
where $S$ is an $SU(2)\;$ $\;2\times 2$ matrix and $\tau_i$ are the
three Pauli matrices. One can immediatelly check that $O_{ij}$ are real
orthogonal 3-parameter matrices with
$O_{ij}O_{kj}=\delta_{ik}$ and $O_{ij}O_{ik}=\delta_{jk}$, as it
should be.

When one rotates the space putting $n_i^\prime=n_jO_{ji}$ the
$2\times 2$ matrix standing in the left upper corner of the ansatz
\ur{leftupper} can be written as

\[
\exp \left[i(\mbox{\boldmath{$n^\prime$}}
\cdot\mbox{\boldmath{$\tau$}}) P(r)\right]
=\cos P(r)+ i(\mbox{\boldmath{$n^\prime$}}
\cdot\mbox{\boldmath{$\tau$}})\sin P(r)
\]
\beq
=S\left[\cos P(r)+ i(\mbox{\boldmath{$n$}}
\cdot\mbox{\boldmath{$\tau$}})\sin P(r)\right]S^\dagger.
\la{cossin}\eeq
Therefore, if one consideres the hedgehog ansatz \ur{leftupper}
rotated {\em both} in flavor and usual spaces, the latter can be
completely absorbed into the former one:

\beq
R\:U_{{\rm cl}}(O\vec{x})\:R^\dagger
=\tilde R\:U_{{\rm cl}}(\vec{x})\:\tilde R^\dagger
\la{equivrot}\eeq
with

\beq
\tilde R= R\left(\begin{array}{cc}S&\begin{array}{c}0\\0\end{array}\\
\begin{array}{cc}0 &\;\;0 \end{array}& 1\end{array}\right).
\la{absorb}\eeq

For that reason it is sufficient to consider rotations only in the
flavor space. Hence there are 3 zero rotational modes in the $SU(2)$
and 8-1=7 in the $SU(3)$ flavor case. The rotation of the form
$R=\exp(i\alpha\lambda^8)$ commutes with the left-upper-corner
ansatz and therefore does not correspond to any zero mode. This will
have important consequences in getting the correct spectrum of
hyperons.

The general strategy is to consider a slowly rotating ansatz

\beq
\tilde U(\vec{x},t)=R(t)\:U_{{\rm cl}}(\vec{x})\:R^\dagger (t)
\la{rotating}\eeq
and to expand the energy of the bound-state level and of the negative
Dirac continuum in `right' ($\Omega_A$) and `left'
($\tilde\Omega_A$) angular velocities

\beq
\Omega_A=-i\Tr(R^\dagger \dot R\lambda^A),\;\;\;\;\;
\tilde\Omega_A=-i\Tr(\dot R R^\dagger\lambda^A),\;\;\;\;\;
\Omega^2=\tilde\Omega^2=2\Tr\dot R^\dagger\dot R.
\la{angvel}\eeq

Taking into account only the lowest powers in the time derivatives
of the rotation matrix $R(t)$ one gets \cite{DPP2,B} the following
form of the rotation lagrangian:

\beq
L^{{\rm rot}} = \frac{1}{2}I_{AB}\Omega_A\Omega_B
-\frac{N_c}{2\sqrt{3}}\Omega_8.
\la{Lrot}
\eeq
Here $I_{AB}$ is the $SU(3)$ tensor of the moments of inertia,
\beq
I_{AB}=\frac{N_c}{4}\int\frac{d\omega}{2\pi}\:\Tr\left(
\frac{1}{\omega+iH}\lambda^A \frac{1}{\omega+iH}\lambda^B\right)
\la{IAB}\eeq
where the $\omega$ integration contour should be drawn {\em above}
the bound-state energy $E_{{\rm level}}$ to incorporate the `valence'
quarks.

The appearance of a linear term in $\Omega_8 $ is an important
consequence of the presence of an extra bound-state level emerging from
the upper Dirac continuum, which fixes the baryon charge to be unity.
In the Skyrme model this linear term arises from the Wess-Zumino
term.$\,$\cite{Gua} For simplicity we have written unregularized
moments of inertia, though \Eq{IAB} should be regularized in some way,
see {\em e.g.} \cite{B}

Owing to the left-upper-corner ansatz for the static soliton
the tensor $I_{AB}$ is diagonal and depends
on two moments of inertia, $I_{1,2}$:

\beq
I_{AB}=\left\{ \begin{array}{cc} I_1\:\delta_{AB},\;&\;A,B=1,2,3,\\
            I_2\:\delta_{AB},\;&\;A,B=4,5,6,7,\\ 0,\;&\;A,B=8.
\end{array}\right.
\eeq
Therefore, the rotational lagrangian \ur{Lrot} can be rewritten as

\beq
L^{{\rm rot}}=\frac{I_1}{2} \sum_{A=1}^{3}\Omega_A^2  +
\frac{I_2}{2} \sum_{A=4}^{7}\Omega_A^2 -\frac{N_c}{2\sqrt{3}}\Omega_8.
\la{rotlagr}
\eeq

To quantize this rotational Lagrangian  one can use the canonical
quantization procedure, same as in the Skyrme
model.$\,$\cite{Gua,DP7,MNP,Chemtob,JW}
Introducing eight angular momenta canonically conjugate to `right'
angular velocities $\Omega_A$,

\beq J_A=\frac{\partial L^{{\rm rot}}}{\partial \Omega_A},
\la{moments}\eeq
and writing the hamiltonian as
\beq
H^{{\rm rot}}=\Omega_A J_A-L^{{\rm rot}}
\la{rothamgen}\eeq
one gets
\beq
H^{{\rm rot}}=\frac{1}{2I_1}\sum_{A=1}^3
J_A^2+\frac{1}{2I_2}\sum_{A=4}^7 J_A^2
\la{Ham1}\eeq
with the additional quantization prescription following from \Eq{moments},

\beq
J_8=-\frac{N_c}{2\surd{3}}=-\frac{\surd{3}}{2}.
\la{quantiz}\eeq

In the Skyrme model this quantization rule follows from the Wess-Zumino
term. In our approach it arises from filling in the
bound-state level, i.e. from the `valence' quarks. It is known to
lead to the selection rule: not all possible spin and $SU(3)$
multiplets are allowed as rotational excitations of the $SU(2)$
hedgehog. \Eq{quantiz} means that only those $SU(3)$ multiplets are
allowed which contain particles with hypercharge $Y=1$; if the number
of particles with $Y=1$ is denoted as $2J+1$, the spin of the allowed
$SU(3)$ multiplet is equal to $J$.

Therefore, the lowest allowed $SU(3)$ multiplets are:

\begin{itemize}
\item octet with spin 1/2 (since there are {\em two} baryons in the octet
with $Y=1$, the $N$)
\item decuplet with spin 3/2 (since there are {\em four} baryons in the
decuplet with $Y=1$, the $\Delta$)
\item anti-decuplet with spin 1/2 (since there are {\em two} baryons in the
anti-decuplet with $Y=1$, the $N^*$)
\end{itemize}
The next are 27-plets with spin $1/2$ and $3/2$ but we do not
consider them here.

We see that the lowest two rotational excitations are exactly the lowest
baryon multiplets existing in reality. The third predicted multiplet,
the anti-decuplet, contains exotic baryons which cannot be made of
three quarks, most notably an exotic $Z^+$ baryon having spin 1/2,
isospin 0 and strangeness +1. A detailed study of the anti-decuplet
performed recently in \cite{DPPol} predicts that such a baryon
can have a mass as low as 1530 MeV and be very narrow. Several
experimental searches of this exotic baryon are now under way.

It is easy to derive the splittings between the centers of the
multiplets listed above. For the representation $(p,q)$ of the $SU(3)$
group one has

\beq
\sum\limits_{A=1}^8 J_A^2 = \frac13[p^2+q^2+pq + 3(p+q)],
\eeq
therefore the eigenvalues of the rotational hamiltonian
\ur{Ham1}) are

\beq
E_{(p,q)}^{{\rm rot}} = \frac1{6I_2} [p^2+q^2+pq + 3(p+q)]
+\left(\frac1{2I_1} - \frac1{2I_2} \right) J(J+1)
- \frac{3}{8I_2}.
\la{rotenerg}\eeq
We have the following three lowest rotational excitations:

\bea
(p,q)=(1,1), & J=1/2:&\; \mbox{octet, spin 1/2}, \\
(p,q)=(3,0), & J=3/2:&\; \mbox{decuplet, spin 3/2}, \\
(p,q)=(0,3), & J=1/2:&\; \mbox{anti-decuplet, spin 1/2} \, .
\eea
The splittings between the centers of these multiplets are determined
by the moments of inertia $I_{1,2}$:

\beq
\Delta_{10-8} =
E_{(3,0)}^{{\rm rot}} - E_{(1,1)}^{{\rm rot}} =\frac{3}{2I_1},
\la{octdecspl}\eeq
\beq
\Delta_{{\overline{10}} - 8} =
E_{(0,3)}^{{\rm rot}} - E_{(1,1)}^{{\rm rot}} =\frac{3}{2I_2}.
\la{oct-antidecspl}\eeq

The appropriate rotational wave functions describing
members of these multiplets are given by Wigner finite-rotation
functions $D^{8,10,{\overline{10}}}(R)$.$\,$\cite{B,DPPol}

When dealing with the flavor $SU(3)$ case neglecting the strange
quark mass $m_s$ is an over-simplification. In fact, it is easy to
incorporate $m_s\neq 0$ in the first order. As a result one gets
very reasonable splittings inside the $SU(3)$ multiplets, as well as
mass corrections to different observables.$\,$\cite{B,Review,DPPol}

In general, the idea that all light baryons are rotational excitations
of one object, the `classical' nucleon, leads to numerous relations
between properties of the members of octet and decuplet, which follow
purely from symmetry considerations and which are all satisfied up
to a few percent in nature. The $SU(3)$ symmetry by itself says
nothing about the relation between different multiplets.
Probably the most spectacular is the Guadagnini formula
\cite{Gua} which relates splittings inside the decuplet with those in
the octet,

\beq
8(m_{\Xi^*}+m_N)+3m_\Sigma=11m_\Lambda+8m_{\Sigma^*},
\la{Guad}\eeq
which is satisfied with better than one-percent accuracy!

\subsection{Case of two flavors}

If one is interested in baryons predominantly `made of' $u,d$ quarks,
the flavor group is $SU(2)$ and the quantization of rotations is
more simple.

In this case the rotational lagrangian is just

\beq
L^{{\rm rot}}=\frac{I_1}{2} \sum_{i=1}^{3}\Omega_i^2
=\frac{I_1}{2} \sum_{A=1}^{3}\tilde\Omega_A^2,
\la{rotlagr2}
\eeq
where the `right' ($\Omega_i$) and `left' ($\tilde\Omega_A$) angular
velocities are defined by \Eq{angvel}. This is the lagrangian for the
spherical top: the two sets of angular velocities have the meaning
of those in the `lab frame' and `body fixed frame'. The quantization
of the spherical top is well known from quantum mechanics. One has
to introduce two sets of angular momenta, $S_i$ (canonically conjugate
to $\Omega_i$) and $T_A$ (conjugate to $\tilde \Omega_A$). Both sets
of operators act on the coordinates of the spherical top, say,
the Euler angles. It will be more convenient for us to say that
the coordinates of the spherical top are just the entries of the
unitary matrix $R$ defining its finite-angle rotation.\cite{DPP2}

The angular momenta operators $S_i,\; T_A$ act on $R$ as
generators of right (left) multiplication,

\beq
e^{i(\alpha S)}Re^{-i(\alpha S)} = R\:e^{i(\alpha\sigma)},
\la{S}\eeq
\beq
e^{i(\alpha T)}Re^{-i(\alpha T)} = e^{-i(\alpha\tau)}\:R,
\la{T}\eeq
and satisfy the commutation relations

\[
[T_A,T_B]=i\epsilon_{ABC}T_C ,\;\;\;\;\;[S_i,S_j]=i\epsilon_{ijk}S_k,
\]
\beq
[T_A,S_i]=0,\;\;\;\;\;\;(T_A)^2=(S_i)^2.
\la{comm}\eeq
A realization of these operators is

\[
S_i=R_{pk}\left(\frac{\sigma_i}{2}\right)_{kq}
\frac{\partial}{\partial R_{pq}},
\]
\beq
T_A=-\left(\frac{\tau_A}{2}\right)_{pk}R_{kq}
\frac{\partial}{\partial R_{pq}}.
\la{realiz}\eeq
The rotational hamiltonian is

\beq
H^{{\rm rot}}=\Omega_i S_i-L^{{\rm rot}}=\tilde\Omega_A T_A-L^{{\rm rot}}
=\frac{S_i^2}{2I_1} =\frac{T_A^2}{2I_1}.
\la{rotham2}\eeq

Comparing the definition of the generators \urs{S}{T} with the
ansatz \ur{rotating} we see that $T_A$ is the flavor (here: isospin)
operator and $S_i$ is the spin operator, since the former acts to the
left from $R$ and the latter acts to the right.

The normalized eigenfunctions of the mutually commuting operators
$S_3,\;T_3$ and $S^2=T^2$ with eigenvalues $S_3,\;T_3$ and
$S(S+1)=T(T+1)$ are \cite{DPP2}

\beq
\Psi^{(S=T)}_{T_3S_3}(R)=\sqrt{2S+1}(-1)^{T+T_3}
D^{(S=T)}_{-T_3S_3}(R)
\la{eigenfu}\eeq
where $D(R)$ are Wigner finite-rotation matrices. For example, in the
$S=T=1/2$ representation $D^{1/2}_{pq}(R)=R_{pq}$, {\em i.e.} coincides
with the unitary matrix $R$ itself.

The rotational energy is thus

\beq
E^{{\rm rot}}=\frac{S(S+1)}{2I_1} =\frac{T(T+1)}{2I_1}
\la{roten2}\eeq
and is $(2S+1)^2=(2T+1)^2$-fold degenerate. The wave functions
\ur{eigenfu} describe at $S=T=1/2$ four nucleon states (proton,
neutron, spin up, spin down) and at $S=T=3/2$ the sixteen
$\Delta$-resonance states, the splitting between them being

\beq
m_{\Delta}-m_N=\frac{3}{2I_1} = O(N_c^{-1})
\la{split2}\eeq
(coinciding in fact with the splitting between the centers of decuplet
and octet in the more general $SU(3)$ case, see \Eq{octdecspl}).

It is remarkable that the nucleon and its lowest excitation, the
$\Delta$, fits into this spin-equal-isospin scheme, following from
the quantization of the hedgehog rotation. Moreover,
since $N$ and $\Delta$ are, in this approach, just different
rotational states of the same object, the `classical nucleon',
there are certain relations between their properties. These relations
are identical to those found first in the Skyrme model \cite{ANW}
since they follow from symmetry considerations only and do not depend
on concrete dynamics which is of course different in the naive
Skyrme model. For example, one gets for the dynamics-independent ratio
of magnetic moments and pion couplings \cite{ANW}

\[
\frac{\mu_{\Delta N}}{\mu_p-\mu_n}=\frac{1}{\surd{2}}\simeq 0.71
\;\;\;\;vs.\;\;\;\;0.70\pm 0.01\;\;({\rm exp.}),
\]
\beq
\frac{g_{\pi N\Delta}}{g_{\pi NN}}=\frac{3}{2}=1.5
\;\;\;\;vs.\;\;\;\;1.5\pm 0.12\;\;({\rm exp.}).
\la{rels2}\eeq

We should mention that there might be interesting implications of
the `baryons as rotating solitons' idea to nuclear physics. The
low-energy interactions between nucleons can be viewed as interactions
between spherical tops depending on their relative orientation
$R_1R_2^\dagger$ in the spin-isospin spaces.\cite{D3,DM} It leads
to an elegant description of $NN$ and $N\Delta$ interactions in a
unified fashion, and it would be very interesting to check its
experimental consequences (as far as we know this has not been done
yet). A nuclear medium is then a medium of interacting quantum
spherical tops with extremely anisotropic interactions depending
on relative orientations of the tops both in the spin-isospin and
in ordinary spaces.

This unconventional point of view is supported by the
observation \cite{D3} that one can get the correct
value of the so-called symmetry energy of the nucleus,
$(25\:{\rm MeV})\cdot(N-Z)^2/A$, the numerical coefficient
appearing as $1/8I_1\simeq 25\:{\rm MeV}$ where $I_1$ is the $SU(2)$
moment of inertia; from the $\Delta - N$ splitting \ur{split2} one
finds $I_1\simeq (200\;{\rm MeV})^{-1}$. We do not know whether the
language of spherical tops is fruitful to describe ordinary nuclear
matter (probably it is but nobody tried), however it is certainly
useful to address new questions, for example whether nuclear matter at
high densities can be in a strongly correlated antiferromagnet-type
phase.$\,$\cite{DM}

Finally, let us ask what the next rotational excitations could be?
If one restricts oneself to only two flavors, the next state should
be a (5/2, 5/2) resonance; in the three-flavor case the next
rotational excitation is the anti-decuplet with spin $1/2$, see
subsection 6.1.  Why do not we have any clear signal of the exotic
(5/2, 5/2) resonance?  The reason is that the angular momentum $J=5/2$
is numerically comparable to $N_c=3$. Rotations with $J\approx N_c$
cannot be considered as slow: the centrifugal forces deform
considerably the spherically-symmetrical profile of the soliton
field;$\,$\cite{DP6,BlR} simultaneously at $J\approx N_c$ the radiation
of pions by the rotating body makes the total width of the state
comparable to its mass.$\,$\cite{DP6,Dorey} In order to survive
strong pion radiation the rotating chiral solitons with $J\ge N_c$ have
to stretch into cigar-like objects; such states lie on linear Regge
trajectories with the slope \cite{DP6} $\alpha^\prime\approx
1/8\pi^2F_\pi^2\approx 1.45\;{\rm GeV}^{-2}$. One cannot exclude
an intriguing possibility that all large-$J$ hadron states lying on
Regge trajectories are actually rotating chiral solitons.$\,$\cite{DP6}

The situation might be somewhat different in the {\it three}-flavor
case. First, the rotation is, roughly speaking, distributed among more
axes in flavor space, hence individual angular velocities are not
necessarily as large as when we consider the two-flavor case with
$J=5/2$. Actually, the $SU(2)$ baryons with $J=5/2$ belong to a very
high multiplet from the $SU(3)$ point of view. Second, the radiation by
the soliton includes now $K$ and $\eta$ mesons which are substantially
heavier than pions, and hence such radiation is suppressed. Actually,
the anti-decuplet seems to have moderate widths \cite{DPPol} and it is
worthwhile to search for the predicted exotic states.

\section{Some applications}

There exists by now a rather vast literature studying baryon observables
in the Chiral Quark-Soliton Model. Baryon formfactors (electric,
magnetic and axial), mass splittings, the nucleon sigma term, magnetic
moments, weak decay constants, tensor charges and many other
characteristics of nucleons and hyperons have been calculated in the
model. We address the reader to an extensive review \cite{Review} on
these matters.

Here we would like to point out several developments of the Chiral
Quark-Soliton Model interesting from the theoretical point of view.
The list below is, of course, very subjective.

The study of the spin content of the nucleon in the model has been
pioneered by Wakamatsu and Yoshiki.$\,$\cite{WY} They showed that the
fraction of the nucleon spin carried by the spin of quarks is
about 50\% (and could be made less): the rest is carried
by the interquark orbital moment, the Dirac sea contribution to it
being quite essential.

An important question is $1/N_c$ corrections to baryon observables.
These can be classified in two groups: one comes from
meson loops and is therefore accompanied by an additional small
factor $\sim 1/8\pi^2$, the second arises from a more accurate
account for the quantization of the zero rotational modes. The
corrections of the second type are not accompanied by small loop
factors, and may be quite substantial: after all in the real world
$N_c=3$ so a 30\% correction is not so small. Such corrections for
certain quantities have been fished out in Refs.(63,64)
in the two-flavor case and in Ref.(65) for three flavors.
These corrections work in a welcome direction: they lower the
fraction of nucleon spin carried by quark spins and increase the
flavor non-singlet axial constants.

Recent applications of the model are to parton distributions in
nucleon, including the so-called skewed distributions, and to the wave
function of the nucleon on the light cone. These topics deserve
special sections.

\section{Nucleon structure functions}

The distribution of quarks, antiquarks and gluons, as measured
in deep inelastic scattering of leptons, provides us probably with the
largest portion of quantitative information about strong interactions.
Until recently only the {\em evolution} of the structure functions
from a high value of the momentum transfer $Q^2$ to even higher values
has been successfully compared with the data. This is the field of
perturbative QCD, and its success has been, historically, essential in
establishing the validity of the QCD itself. However, the initial
conditions for this evolution, namely the leading-twist distributions
at a relatively low normalization point, belong to the field of
non-perturbative QCD. If we want to understand the vast amount of data
on unpolarized and polarized structure functions we have to go into
non-perturbative physics.

The Chiral Quark-Soliton Model presents a non-perturbative approach
to the nucleons, and it is worthwhile looking into the parton
distributions it predicts. Contrary to several models of nucleons
on the market today, it is a relativistic field-theoretic model.
This circumstance becomes of crucial importance when one deals with
parton distributions. It is only with a relativistic field-theoretic
model one can preserve general properties of parton distributions such
as

\begin{itemize}
\item  relativistic invariance,
\item  positivity of parton distributions,
\item  partonic sum rules which hold in full QCD.
\end{itemize}

There are two seemingly different ways to define parton distributions.
The first, which we would call the Fritsch--Gell-Mann definition, is
a nucleon matrix element of quark bilinears with a light-cone
separation between the quark $\psi$ and $\bar\psi$ operators.
According to the second, which we would call the Feynman--Bjorken
definition, parton distributions are given by the number of partons
carrying a fraction $x$ (the Bjorken variable) of the nucleon
momentum in the nucleon infinite-momentum frame. See Feynman's book
\cite{F} for the discussion of both definitions. In perturbative QCD
only the Fritsch--Gell-Mann definition has been exploited as it is very
difficult to write down the nucleon wave function in the
infinite-momentum frame, which is necessary for the Feynman--Bjorken
definition.

Despite the apparent difference in wording, it has been shown for the
first time, within the field-theoretic Chiral Quark-Soliton Model,
that the two definitions are, in fact, equivalent and lead to
identical working formulae for computing parton distributions:
in Ref.(67) the first definition has been adopted while
in Ref.(68) the second was used. The deep reason for that
equivalence is that the main hypothesis of the Feynman--Bjorken
parton model is satisfied in the Chiral Quark-Soliton Model,
namely that parton transverse momenta do not grow with $Q^2$.

Let us point out some key findings of Refs.(67,68).

\subsection{Classification of quark distributions in $N_c$}

Since the nucleon mass is $O(N_c)$ all parton distributions are
actually functions of $xN_c$. Combining this fact with the known
large-$N_c$ behavior of the integrals of the distributions over
$x$ one infers that all distributions can be divided into `large'
and `small'. The `large' distributions are, for example, the
unpolarized singlet and polarized isovector distributions, which are of
the form

\beq
D^{{\rm large}}(x)\sim N_c^2f(xN_c),
\la{large}\eeq
where $f(y)$ is a stable function in the large-$N_c$ limit. On the
contrary, the polarized singlet and unpolarized isovector distributions
give an example of `small' distributions, having the form

\beq
D^{{\rm small}}(x)\sim N_cf(xN_c).
\la{small}\eeq
One, indeed, observes in experiment that `large' distributions are
substantially larger than the `small' ones.

It should be mentioned that in the large-$N_c$ limit there
exist certain strict inequalities between various polarized
and unpolarized distributions.$\,$\cite{PPner}

\subsection{Antiquark distributions}

In the academic limit of a very weak mean pion field in the nucleon
the Dirac continuum reduces to the free one (and should be subtracted
to zero) while the bound-state level joins the upper Dirac continuum.
In such a limit there are no antiquarks, while the distribution of
quarks becomes $q(x)=N_c^2\delta(xN_c-1)$. In reality there is a
non-trivial mean pion field which ({\em i}) creates a bound-state
level, ({\em ii}) distorts the negative-energy Dirac continuum. As the
result, the above $\delta$-function is smeared significantly, and a
non-zero antiquark distribution appears.

An inevitable consequence of the relativistic invariance is that the
bound-state level produces a {\em negative} contribution to the
antiquark distribution.$\,$\footnote{This is also true for any nucleon
model with valence quarks, for example for any variant of bag models.
Bag models are essentially non-relativistic, so they fail to resolve
this paradox. In order to cure it, one has to take into account
contributions to parton distributions from {\em all} degrees of freedom
involved in binding the quarks in the nucleon. That can be consistently
done only in a relativistic field-theoretic model, like the one
under consideration.} The antiquark distribution becomes positive
only when one includes the contribution of the Dirac continuum.
Numerically, the antiquark distribution appears to be sizeable
even at a low normalization point, in accordance with phenomenology.

\subsection{Sum rules}

The general sum rules holding in full QCD are automatically satisfied
in the Chiral Quark-Soliton Model: in \cite{SF1,SF2} the validity
of the baryon number, isospin, total momentum and Bjorken sum rules
has been checked. In fact, it is for the first time that nucleon
parton distributions at a low normalization point have been
consistently calculated in a relativistic model preserving all general
properties.

\subsection{Smallness of the gluon distribution}

As many times stressed in this paper, the whole approach of the
Chiral Quark-Soliton Model is based on the smallness of the algebraic
parameter $(M\bar\rho)^2$ where $1/\bar\rho$ is the UV cutoff off the
low-momenta theory, for example, the average size of instantons in the
vacuum.  This $\bar\rho$ is the size of  the constituent quark, while
the size of the nucleon is, parametrically, $1/M$. Computing parton
distributions in the model one is restricted to momenta $k\sim M
\approx 350\:{\rm MeV}$, so that the internal structure of the
constituent quarks remains unresolved.  There are no gluons in the
nucleon at this resolution scale; indeed, the momentum sum rule is
satisfied with quarks and antiquarks only.

However, when one moves to the resolution scale of 600 MeV or
higher, the constituent quarks cease to be point-like, and that is
at this scale that a non-zero gluon distribution emerges. Having
a microscopic theory of how quarks get their dynamical masses
one can, in principle, compute the non-perturbative gluon distribution
in the constituent quarks. What can be said on general
grounds is that the fraction of momentum carried by gluons is of
the order of $(M\bar\rho)^2\approx 1/3$, which seems to be the
correct portion of gluons at a low normalization point of about 600
MeV where the normal perturbative evolution sets in.

\subsection{Comparison with phenomenology}

There are several parametrizations of the nucleon parton
distributions at a relatively low normalization point, which,
after their perturbative evolution to higher momentum transfer
$Q^2$, fit well the numerous data on deep inelastic scattering.
The parametrization most convenient for our purpose is
that of Gl\"uck, Reya {\em et al.} \cite{GR1,GR2} who pushed the
normalization point for their distributions to as low as 600 MeV,
starting from the perturbative side. In Refs.(67,68) parton
distributions following from the Chiral Quark-Soliton Model have been
compared with those of Refs.(70,71), see Fig. 5. There seems to
be a good qualitative agreement though the constituent quark and
antiquark distributions appear to be systematically `harder' than those
of.$\,$\cite{GR1,GR2} This deviation is to be expected since the
calculations refer to even lower normalization point where the
structure of the constituent quarks themselves is not been taken
into account yet, see above.

\begin{figure}
\centerline{\psfig{figure=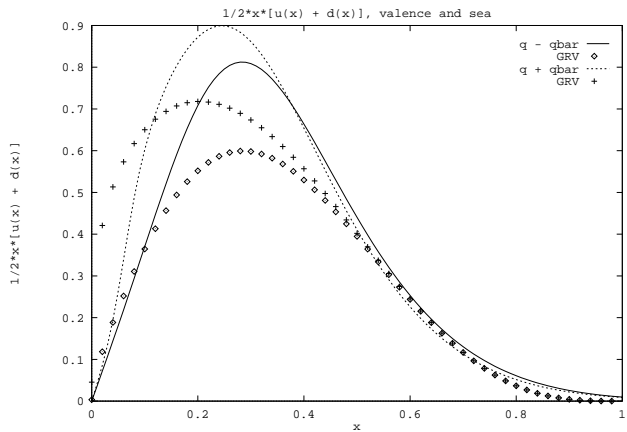,height=6.0cm}}
\caption
{Singlet unpolarized structure functions
$x\{[u(x)+d(x)]\pm[\bar u(x)+\bar d(x)]\}/2$ from
Ref.(68) compared to the phenomenological parametrization
of Ref.(70). The dashed line corresponds to the sum and the solid
line corresponds to the difference of quark and antiquark
distributions.}
\end{figure}

\subsection{Flavor asymmetry of the sea}

An important feature of the model is the large flavor asymmetry
of the sea. Namely, isovector {\em antiquark} distributions are not
small, especially the polarized $\Delta\bar u-\Delta\bar d$
distribution which is of the leading order in $N_c$, while the
unpolarized $\bar u - \bar d$ distribution is of the subleading order.
Phenomenologically, it is a welcome feature but which is hard to
accommodate in other nucleon models. In the Chiral Quark-Soliton
Model it arises in a most natural way since the nucleon is bound by
the isospin 1 pion field.

In Fig. 6 we present the calculated $\bar u - \bar d$
distribution \cite{flavor2} as compared to the experimental data.
See \cite{flavor} for the latest discussion of the consequences of
large flavor asymmetry of the sea on the description of the data.

Finally, we mention that the Chiral Quark-Soliton Model has been
used to make predictions on the transversity spin
distributions.$\,$\cite{PP}

\begin{figure}
\centerline{\psfig{figure=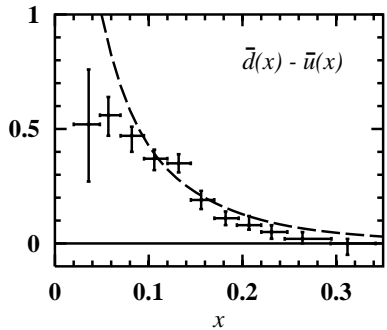,height=5.0cm}}
\caption
{The antiquark asymmetry $\bar d(x) - \bar u(x)$ in the proton
computed in Ref.(72) and perturbatively evolved to $Q = 7.35\,{\rm
GeV}$, as compared to the FNAL E866 data of Ref.(75).}
\end{figure}

\section{Skewed parton distributions}

Recently a new type of parton distributions
\cite{Bartels}$\!^-\!$\cite{CFS} has attracted considerable
interest, the so-called skewed (also called non-forward, off-forward,
non-diagonal) parton distributions (the SPD's) which are
generalizations of ({\em i}) usual parton distributions, ({\em ii})
distribution amplitudes and ({\em iii}) of the elastic nucleon form
factors, in the case of vacuum quantum numbers in the $t$ channel. At
the same time there is a wide range of processes where one probes
skewed densities which do not have a diagonal analog.$\,$\cite{CFS}

The SPD's are not accessible in standard inclusive measurements.
However, they can be measured in diffractive photoproduction of the
$Z$ boson,$\,$\cite{Bartels} in diffractive electroproduction of vector
mesons,$\,$\cite{AFS,CFS,Rad2} in deeply-virtual Compton
scattering \cite{Ji1,Rad,Rad2} and in hard exclusive electroproduction
of mesons at moderate values of Bjorken $x$.$\,$\cite{CFS,Piller}

The number of papers devoted to skewed distributions is rather large by
now and continue to grow rapidly. Computation of the SPD's at a low
normalization point in a realistic model of nucleons is of great
importance. While the usual parton distributions have been measured in
a variety of different experiments and can be confronted
{\em a posteriori} with model calculations, in the case of the
off-forward distributions the situation is opposite: model calculations
of the distributions at a low normalization point are required to
determine the very feasibility of measuring the SPD's.

First model calculations of skewed distributions were performed
in the bag model.$\,$\cite{Ji3} However, bag models
encounter severe problems in applications to parton distributions.
It is usually assumed that the three quarks in the bag give
rise only to quark distributions; however, they produce also
an antiquark distribution with a {\em negative} sign, as an inevitable
consequence of relativistic invariance.
To overcome this problem one needs to add the contribution from forces
which bind the quarks -- in the case of the bag model it is the
mysterious bag surface whose `structure function' will hardly be ever
defined. The Chiral Quark-Soliton Model has no difficulties of that
kind, and skewed distributions have been calculated in the
model.$\,$\cite{PPPBGW} They appear to be {\em qualitatively}
different from those of the bag model. We emphasize that this
calculation is done in a self-consistent, relativistic-invariant
model which ensures the positivity of both quark and antiquark
distributions.

Skewed parton distributions are defined through
non-diagonal matrix elements of the product of quark fields at a
light--cone separation. We shall use the notations of
Ji \cite{Ji1,Ji3} for the two functions depending on three
variables:

\[
\nonumber
\int \frac{d\lambda }{2\pi }e^{i\lambda x}
\langle P^{\prime }|\bar \psi(-\lambda n/2)
\psi(\lambda n/2)|P\rangle = H(x,\xi,\Delta^2)\,
\left(\bar u(P^\prime)\,{\Dirac n}\,u(P)\right)
\]
\beq
+ \frac 1{2{\cal M}_N}\, E(x,\xi,\Delta^2)\,
\left(\bar u(P^\prime)\,i\sigma^{\mu\nu}n_\mu\Delta_\nu\,u(P)\right).
\nonumber
\label{E-H-QCD-2}
\eeq
Here $n_\mu$ is a light-cone vector, $n^2 = 0$, $n\cdot (P+P')=2$,
and $\Delta$ is the four-momentum transfer, $\Delta = P^\prime -P$.
${\cal M}_N$ as usually denotes the nucleon mass, and
$u(P),\,\bar u(P^\prime)$ are the Dirac spinors of the nucleon in
the initial and final states. The off-forward quark distributions,
$H(x,\xi,\Delta^2)$ and $E(x,\xi,\Delta^2)$, are regarded as
functions of the variable $x$, the square of the four-momentum
transfer $\Delta^2$ and of its longitudinal component
$\xi=-(n\cdot \Delta)$

In the forward case, $P=P'$, both $\Delta$ and $\xi$ are zero, and the
second term on the r.h.s. of \Eq{E-H-QCD-2} vanishes. The
function $H$ becomes the usual parton distribution function,

\beq H(x,\xi =0,\Delta^2=0)=q(x) .
\la{forward_limit}
\eeq
On the other hand, taking the first moment of \Eq{E-H-QCD-2} one
reduces the operator on the l.h.s. to the local vector current.
The dependence of $H$ and $E$ on $\xi$ disappears, and the functions reduce
to the usual electric and magnetic form factors of the nucleon,
\bea
\nonumber
\int_{-1}^1 dx \; H(x,\xi,\Delta^2)&=&F_1(\Delta^2), \\
\nonumber
\int_{-1}^1 dx \; E(x,\xi,\Delta^2)&=&F_2(\Delta^2).
\eea

The large $N_c$ limit which is the basis of all calculations in the
Chiral Quark-Soliton Model allows to calculate off-forward
distributions in the region $x,\xi \sim 1/N_c,\quad
\Delta^2 \sim N_c^0$. Also, it introduces a classification
of the skewed distribution in $N_c$: in the leading
order in $N_c$ only isosinglet part of $H(x,\xi ,\Delta ^2)$
and  isovector part of $E(x,\xi ,\Delta ^2)$ appear.
Isovector $H(x,\xi ,\Delta ^2)$ and isosinglet
$E(x,\xi ,\Delta ^2)$ are subleading in $N_c$: they appear
after taking into account the finite angular velocity of the soliton
rotation.

The calculation of the off-forward parton distributions proceeds in much
the same way as that of the usual parton distributions.
A typical off-forward distribution obtained from the Chiral
Quark--Soliton Model is plotted in Fig. 7. Notice the
oscillating character of the distribution with
sharp crossovers near the points  $x = \pm\xi/2$. It can be shown on
general grounds \cite{PPPBGW} that they should be present almost in
any self-consistent field-theoretic model.$\,$\footnote{They are absent
in bag model calculations.$\,$\cite {Ji3}} At small $\Delta$
(that is in the forward limit) these points divide the interval
$-1 \le x \le 1$ into three regions: ({\it i}) $x\ge \xi/2$, where the
skewed distribution $H(x,\xi,\Delta^2)$ is close to the quark parton
distribution, ({\it ii}) $x\le -\xi/2$  where it is close to
the antiquark parton distribution (with a minus sign) and
({\it iii}) the intermediate region where it has a sharp crossover.
The crossover is due to the fact that both quark and
antiquark distributions should be positive. It can be
shown \cite{PPPBGW} that the dependence of the quark mass $M(k)$ on the
quark virtuality converts a jump (with infinite derivative) in  the
skewed distributions into a sharp crossover with the width of the
order $\sim (1/M\bar\rho)^2$.

The fast crossover of $H(x,\xi,\Delta^2)$ at $x=\pm \xi/2$
may have interesting physical implications. For example, it may lead to
a considerable enhancement of the deeply-virtual Compton scattering
cross section:  parametrically it is enhanced by a factor $(\log
M\bar\rho)^2$.

Applications of the Chiral Quark-Soliton Model to {\em polarized}
skewed distributions can be found in Ref.(85). The model
predicts a considerable enhancement of the hard exclusive production
of charged pions \cite{strikfurtmax,MPR} accompanied by a large
azimuthal spin asymmetry.$\,$\cite{strikfurtmax} Also, it appears
possible to obtain certain model-independent relations between
different skewed distributions (based on the large $N_c$ limit), to
calculate the transition $N\to\Delta$ distributions, etc.

\begin{figure}
\centerline{\psfig{figure=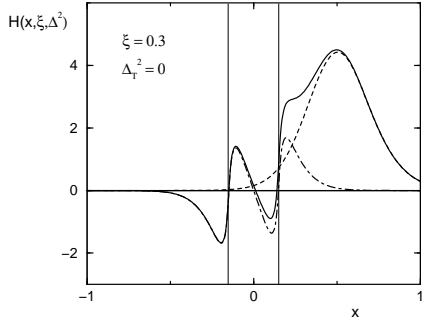,height=5.0cm}}
\caption{Prediction for the isosinglet skewed distribution
$H(x,\xi,\Delta^2)$ for $\Delta_T^2=0$
($\Delta_T^2 \equiv -\Delta^2-\xi^2 M_N^2$) and $\xi=0.3$.
{\em Dashed line}: contribution from the discrete level;
{\em dashed-dotted line}: contribution from the Dirac continuum;
{\em solid line}: the total distribution (sum of the two).
The vertical lines mark the crossover points $x=\pm \xi/2$.}
\end{figure}

\section{Light-cone wave function of the nucleon}

Probably, the most complete information about processes at high
energies involving nucleons can be extracted from its wave
function on the light cone or, else, in the infinite momentum
frame. Using the infinite momentum frame allows one to separate the
wave function of the nucleon from vacuum fluctuations. The wave
function is, generally, a Fock vector which determines amplitudes to
find inside the nucleon three quarks, or three quarks plus
a quark-antiquark pair, or three quarks plus one gluon, etc. The wave
function squared gives the probability to find quarks inside the
nucleon with given fractions of nucleon longitudinal momentum and some
transverse momenta. Integrating it over transverse momenta and
summing up over all possible configurations with one quark
(or antiquark) fixed one arrives to standard parton distributions.

By the wave function in a `narrow sense' $\phi(z_1,z_2,z_3)$
(else called {\em distribution amplitude}) people usually understand
the amplitude to find exactly three quarks inside the nucleon with
given fractions $z_1,z_2,z_3$ of the nucleon longitudinal momentum. It
determines the high-energy asymptotics of exclusive processes such as
the asymptotics of electromagnetic formfactors, the decays $J/\psi\to
N\bar{N}$, etc. It is rather well known from experiment (see,
e.g. \cite{kroll}).

The Chiral Quark-Soliton Model presents a unique possibility
to determine the whole Fock state vector of the nucleon on the
light cone. The nucleon wave function (as well as parton distributions)
are subject to the evolution in perturbation theory, therefore we are
actually speaking about the wave function at a low normalization point.
This work is in progress, so we present some preliminary
results.$\,$\cite{MYNWF}

In the Chiral Quark-Soliton Model it is easy to define the wave
function of the nucleon in the rest frame. Indeed, the nucleon in the
model represents quarks in the Hartree approximation in the
self-consistent pion field. In this approximation the full wave
function is the product of one-particle states,

\beq
|\Phi_N>=\prod_{{\rm occupied}}\int\! d^3x f^{(n)}(x)\Psi^+(x)|0>,
\la{Dirac}
\eeq
where $f^{n}(x)$ are the eigenfunctions of the time-independent Dirac
operator in the external pion field, $Hf^{(n)}=E^{(n)}f^{(n)}$,  with
the hamiltonian \ur{DirHam}. The product goes over all occupied levels,
i.e. over the discrete `valence' level and over the negative-energy
continuum.

Actually we need the wave function not in terms of the Dirac continuum
but rather in terms of quarks and antiquarks. It is well known that the
wave function \ur{Dirac} corresponds to the so-called coherent exponent:

\[
|\Phi_N>=c_{00}\prod_{{\rm color}}\int dk\, F^{\lambda}(k)\,
{\bf a}^{\lambda+}(k_1)
\]
\beq
\times \exp \int dk_1 dk_2\, {\bf a}^{\lambda_1+}(k_1)\,
\Theta_{\lambda_1\lambda_2}(k_1,k_2)\,{\bf b}^{\lambda_2+}(k_2),
\la{coherent}\eeq
where ${\bf a}^+,{\bf b}^+$ are quark and antiquark creation operators.
It represents $N_c$ quarks with the wave function $F^{\lambda}(k)$
and a number of additional quark-antiquark pairs whose
wave functions in the external pion field are
$\Theta_{\lambda_1\lambda_2}(k_1,k_2)$; they can be calculated using
the Feynman Green function at finite time, which can be
constructed from the Dirac equation. As to the single-quark wave
function $F^{\lambda}(k)$ it gets contributions both from the
discrete level and from the Dirac continuum described by the
pair wave function $\Theta_{\lambda_1\lambda_2}(k_1,k_2)$.

We next proceed as follows:
\begin{itemize}
\item
The wave function \ur{coherent} is translated to the infinite
momentum frame. This is possible as the model is
relativistically invariant.
\item
The pair wave functions are calculated
using the technique of interpolating formulae described in
subsection 4.1. It is interesting to note that higher terms of this
expansion give the wave functions with 1,2,$\ldots$ additional pions
on the light cone.$\,$\cite{pionwf,maxchrist} This is to be expected,
as pions are the only agents inducing the interaction in the model.
\item
The result should be averaged with the rotational wave function
defined in section 6.2 to project the general wave function onto a
given rotational state of the soliton -- the nucleon, the $\Delta$,
etc.
\end{itemize}

The three-quark contribution to the (proton, spin up) wave function is:

\beq |\Phi_N,{\rm proton}\uparrow>=
\Phi(z_1,p_{1\perp},z_2,p_{2\perp},z_3,p_{3\perp})\;
|u\!\uparrow\!(1)> |u\!\downarrow\!(2)>
|d\!\uparrow\!(3)>.
\eeq
After integrating over transverse momenta of quarks one gets the
distribution amplitude. The discrete-level contribution to it has the
form:

\[
\phi(z_1,z_2,z_3)= \int d^2x_\perp \left\{
g(z_1,x_\perp) g(z_2,x_\perp)g(z_3,x_\perp)\right.
\]
\beq
+\left. x_\perp^2j(z_2,x_\perp) \left[ 2g(z_1,x_\perp)j(z_3,x_\perp)
-j(z_1,x_\perp)g(z_3,x_\perp)\right]\right\},
\la{nwflevel}
\eeq
where
\bea
g(z,x_\perp)&=&-\frac{\pi^{3/2}}{2}\!\int\! dx_3\exp(-i(z
{\cal M}_N-E_{\rm level}))\!
\left[h(r)-\frac{ix_3}{r}j(r)\right]\!\!,\\
j(z,x_\perp)&=&\frac{\pi^{3/2}}{2}\!\int\! dx_3\exp(-i(z
{\cal M}_N-E_{\rm level})) \left[\frac{i}{r}j(r)\right].
\eea
Here $h(r)$ and $j(r)$ are the wave functions of the discrete
level \ur{K0} and $E_{\rm level}$ is its energy.

Let us note that the second term in curly brackets in \Eq{nwflevel}
introduces an asymmetry of the wave function under the exchange of
$z_i$'s.  This effect is due to the finiteness of $N_c$ and appears as a
result of averaging over nucleon rotational wave function. At
$N_c\rightarrow\infty$ the wave function of the nucleon would be
completely symmetric.

\begin{figure}
\centerline{\psfig{figure=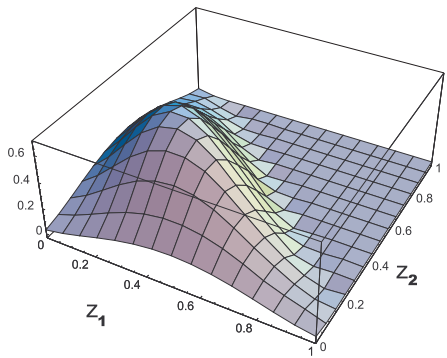,height=5.0cm}}
\caption{Nucleon wave function $\phi(z_1,z_2,1-z_1-z_2)$.
Only the `valence' level contribution is included.}
\end{figure}

The calculated distribution amplitude \ur{nwflevel}
is presented in Fig. 8. Though not fully symmetric under
the interchange of $z_i$'s, it is much closer to the asymptotic
nucleon wave function,
\beq
\phi^{{\rm asympt}}(z_1,z_2,z_3)=120\,z_1z_2z_3\,\delta(z_1+z_2+z_3-1),
\eeq
than the wave function suggested by Chernyak, Ogloblin and Zhitnisky
\cite{COZ} many years ago on the basis of their analysis of the QCD sum
rules.$\,$\footnote{The {\em pion} wave function at a low normalization
point as calculated from the instanton vacuum \cite{pionwf} also
appears to be rather close to its asymptotic form.} It seems to be
supported by the data.$\,$\cite{kroll}

\section{Conclusions}

The Chiral Quark-Soliton Model of baryons is a simple but still a
non-trivial reduction of the full-scale QCD at low energies.
It emphasizes the spontaneous chiral symmetry breaking in QCD,
accompanied by the appearance of the dynamical (or constituent)
quark mass. We prefer the word `dynamical': first, because it is,
indeed, dynamically generated, second, because it is momentum-dependent.

The momentum dependence of the dynamical quark mass $M(k)$ is the
key to understanding why the notion of constituent quarks have worked
so remarkably well over 30 years in hadron physics. The point is, the
scale $\Lambda$ at which the function $M(k)$ falls off appears to be
much larger than $M(0)$; the former determines the size of constituent
quarks while the latter determines the size of hadrons. As a matter of
fact these two distinctive scales come neatly from instantons.

The Chiral Quark-Soliton Model fully exploits the existence of the
two distinctive scales: it is because of them it makes sense to
restrict oneself to just two degrees of freedom in the nucleon problem,
namely, to massless or nearly massless (pseudo) Goldstone pions and to
the constituent quarks with a momentum-dependent dynamical mass
$M(k)$. The scale $\Lambda$ actually plays the role of the physical
ultra-violet cutoff for the low-energy theory; its domain of
applicability is thus limited to the range of momenta
$k\sim M < \Lambda$. This is precisely the domain of interest for
the nucleon binding problem.

A technical tool simplifying considerably the nucleon problem is
the use of the large $N_c$ logic. At large $N_c$ the nucleon is
heavy, and one can speak of the classical self-consistent pion
field binding the $N_c$ valence quarks of the nucleon together.
The classical pion field (the soliton) is found from minimizing the
energy of the bound-state level plus the aggregate energy of the
lower Dirac continuum in a trial pion field. The `valence'
quarks sitting on the bound-state level appear to be strongly
bound by the classical pion field. The system is relativistic.

By quantizing the slow rotations of the soliton field in flavor
and ordinary spaces one gets baryon states which are rotational
excitations of the static `classical nucleon'. The classification
of the rotational excitations depends on the symmetry properties of
the soliton field, but not on the details of dynamics. Taking
the hedgehog ansatz one gets the following lowest baryon
multiplets: octet with spin 1/2, decuplet with spin 3/2 (these are,
indeed, the lowest multiplets observed in nature) and antidecuplet
with spin 1/2. This last multiplet contain baryons with exotic
quantum numbers (in the sense that they cannot be composed of only
three quarks); some of them are predicted to be relatively light
and narrow resonances, and it would be of great interest to search
for such states.

By saying that all lightest baryons are nothing but rotational
excitations of the same object, the `classical nucleon', one gets
many relations between members of baryon multiplets, which are
realized with astonishing accuracy in nature. Especially successful
are predictions which do not depend on dynamical quantities (like
the values of moments of inertia) but follow from symmetry
considerations only, and are therefore shared, {\em e.g.}, by the
Skyrme model. Predictions of the Chiral Quark-Soliton Model which
do depend on concrete dynamics are, in general, also in good
accordance with reality: the typical accuracy for numerous baryon
observables computed in the model is about 15-20\%, coinciding
with the expected theoretical accuracy of the model. To get a
better accuracy one needs a better understanding of the underlying
QCD vacuum and of the resulting effective low-energy theory.

To our knowledge, the Chiral Quark-Soliton Model is today the only
relativistic field-theoretic model of the nucleon, and this advantage
of the model becomes crucial when one turns from static
characteristics of the nucleon to the numerous parton distributions.
It is impossible to get a consistent description of parton
distributions satisfying positivity and sum rules restrictions, without
having a relativistic theory at hand and without taking into account
the complete set of forces which bind quarks together. The leading-twist
parton distributions calculated in the Chiral Quark-Soliton Model refer
to a very low normalization point where the structure of the
constituent quarks is not resolved yet. They seem to be in qualitative
agreement with parametrizations of the DIS data at low $Q^2$ though,
not unnaturally, they appear to be more `hard'. One of the striking
predictions of the model is large flavor asymmetry of antiquark
distributions, which is favored by recent experiments. An even larger
asymmetry is predicted for polarized antiquark distributions, but it
has not been directly measured so far.

The Chiral Quark-Soliton Model has been recently used to calculate
some of the off-diagonal or skewed distribution functions which provide
even a more detailed information on the nucleon structure than the
usual (diagonal) parton distributions. It predicts a drastic
oscillating form of the skewed distribution. Preliminary results
have been obtained for the nucleon wave function on the light cone.

An impressive amount of data on parton distributions has been
collected in the past years, and its amount will grow with new types
of distributions becoming experimentally accessible. We know how to
(perturbatively) evolve parton distributions from high to still higher
values of $Q^2$ but we do not understand how to explain the
initial conditions for that evolution, since it involves truly
nonperturbative physics. This is a real challenge to the theory
of strong interactions. At present the only nonperturbative
relativistic field-theoretic model of QCD at low energies is the
`Nucleons as Chiral Solitons' model. Therefore, we think that
it will play a major role in understanding, describing and predicting
data in the years to come.

\section*{Acknowledgments}

We are grateful to Klaus Goeke for his invariably friendly hospitality
at Bochum University where big portions of the work reviewed here has
been done, and to many other our collaborators, especially to Pavel
Pobylitsa and Maxim Polyakov, for numerous discussions over the past
years of the topics presented here. V.P. acknowledges partial support
by the Russian Foundation for Basic Research, grant RFBR-00-15-96610.


\section*{References}
\begin{thebibliography}{99}

\bibitem{SVZ}
M.A. Shifman, A.I. Vainshtein and V.I. Zakharov, \Journal{\NPB}{147}
{385}{1979}.

\bibitem{Ioffe}
B.L. Ioffe, \Journal{\NPB}{188}{317}{1981}.

\bibitem{GL}
J. Gasser and H. Leutwyler, \Journal{\NPB}{250}{465}{1985}.

\bibitem{BCG}
J. Bijnens, G. Colangelo and J. Gasser, \Journal{\NPB}{427}{427}{1994}.

\bibitem{Knecht}
M. Knecht, B. Moussallam, J. Stern and N.H. Fuchs,
\Journal{\NPB}{457}{513}{1995}.

\bibitem{WZ}
J. Wess and B. Zumino, \Journal{\PLB}{37}{95}{1971}.

\bibitem{DE}
D. Diakonov and M. Eides, {\it Sov. Phys. JETP Lett.} {\bf 38}, 433
(1983).

\bibitem{Dhar}
A. Dhar and S. Wadia, \Journal{\PRL}{52}{959}{1984};\\
A. Dhar, R. Shankar and S. Wadia, \Journal{\PRD}{31}{3256}{1985}.

\bibitem{W}
E. Witten, \Journal{\NPB}{223}{422}{1983}.

\bibitem{Skyrme}
T.H.R. Skyrme, {\it Nucl. Phys.} {\bf 31}, 556 (1962).

\bibitem{ANW}
G. Adkins, C. Nappi and E. Witten, \Journal{\NPB}{228}{552}{1983}.

\bibitem{DP3}
D. Diakonov and V. Petrov, \Journal{\NPB}{272}{457}{1986}.

\bibitem{DP4}
D. Diakonov and V. Petrov in {\it Hadron Matter under Extreme
Conditions}, eds. G. Zinoviev and V. Shelest (Naukova dumka, Kiev,
1986) p.192; \\
{\it Spontaneous Breaking of Chiral Symmetry in the Instanton
Vacuum}, preprint LNPI-1153 (1986).

\bibitem{DP1}
D. Diakonov and V. Petrov, \Journal{\NPB}{245}{259}{1984}.

\bibitem{DP2}
D. Diakonov and V. Petrov, \Journal{\PLB}{147}{351}{1984},\\
{\it Sov. Phys. JETP} {\bf 62}, 204 (1985); {\it ibid.} {\bf 62}, 431
(1985).

\bibitem{tH}
G. 't Hooft, \Journal{\PRD}{14}{3432}{1976}.

\bibitem{BC}
T. Banks and A. Casher, \Journal{\NPB}{169}{103}{1980}.

\bibitem{DPW}
D. Diakonov, M. Polyakov and C. Weiss, \Journal{\NPB}{461}{539}{1996};
{\tt hep-ph/9510232}.

\bibitem{D1}
D. Diakonov in {\it Selected Topics in Non-Perturbative QCD}, eds.
A. Di Giacomo and D. Diakonov (Amsterdam, 1996) p. 397,
{\tt hep-ph/9602375}.

\bibitem{NJL}
Y. Nambu and G. Jona-Lasinio, {\it Phys. Rev.} {\bf 122}, 345 (1961);
{\it ibid.} {\bf 124}, 246 (1961).

\bibitem{D4}
D. Diakonov in {\it Advanced School on Non-Perturbative Quantum Field
Physics}, eds. M. Asorey and A. Dobado (World Scientific, Singapore,
1998), p. 1, {\tt hep-ph/9802298}.

\bibitem{MG}
A. Manohar and H. Georgi, \Journal{\NPB}{234}{189}{1984}.

\bibitem{Volkov} M.K. Volkov and D. Ebert, {\it Sov. J. Nucl. Phys.}
{\bf 36}, 736 (1982);
\Journal{\ZPC}{16}{205}{1983};\\
M.K. Volkov, {\it Ann. Phys.} {\bf 157}, 282 (1984);
{\it Sov. J. Part. Nucl.} {\bf 17}, 186 (1986).

\bibitem{DPP2}
D. Diakonov, V. Petrov and P. Pobylitsa, in {\it Elementary Particle
Physics}, Proc. 21st PNPI Winter School (Leningrad, 1986) p.158; \\
D. Diakonov, V. Petrov and P. Pobylitsa, \Journal{\NPB}{306}{809}{1988}.

\bibitem{DPY}
D. Diakonov, V. Petrov and A. Yung, \Journal{\PLB}{130}{385}{1983};
{\it Sov. J. Nucl. Phys.} {\bf 39}, 150 (1984).

\bibitem{Z}
J. \.Zuk, \Journal{\ZPC}{29}{303}{1985}.

\bibitem{PV}
M. Prasza\l ovicz and G. Valencia, \Journal{\NPB}{341}{27}{1990}.

\bibitem{PolVer}
M. Polyakov and V. Vereschagin, \Journal{\PRD}{54}{1112}{1996}.

\bibitem{DP7}
D. Diakonov and V. Petrov, {\it Baryons as Solitons}, preprint
LNPI-967 (1984), published in {\it Elementary Particles},
(Energoatomizdat, Moscow, 1985) vol.2, p.50.

\bibitem{DHF}
E. D'Hoker and E. Farhi, \Journal{\NPB}{241}{109}{1984}.

\bibitem{MKW}
R. MacKenzie and F. Wilczek, \Journal{\PRD}{30}{2194}{1984}.

\bibitem{NS}
A.J. Niemi and G.W. Semenoff, {\it Phys. Rept.} {\bf 135}, 99 (1986).

\bibitem{Pol}
A.M. Polyakov, \Journal{\NPB}{120}{429}{1977}.

\bibitem{DP9}
D. Diakonov and V. Petrov, {\it Phys. Scripta} {\bf 61}, 536 (2000),
{\tt hep-lat/9810037}.

\bibitem{Wup}
G. Bali, C. Schlichter and K. Schilling,
\Journal{\PRD}{51}{5165}{1995}.

\bibitem{Gr}
V.N. Gribov, {\it Phys. Scripta} {\bf 15}, 164 (1987);
{\tt hep-ph/9807224}.

\bibitem{DP8}
D. Diakonov and V. Petrov, {\it Sov. Phys. JETP Lett.} {\bf 43}, 57
(1986);  \\
D. Diakonov, in: {\em Skyrmions and Anomalies}, eds. M. Je\.zabek and
M. Prasza\l owicz, (World Scientific, Singapore, 1987) p.27.

\bibitem{KRS}
S. Kahana, G. Ripka and V. Soni, \Journal{\NPA}{415}{351}{1984};\\
S.Kahana and G.Ripka, \Journal{\NPA}{429}{462}{1984}.

\bibitem{BB}
M.S. Birse and M.K. Banerjee, \Journal{\PLB}{136}{284}{1984}.

\bibitem{Soni}
V. Soni, \Journal{\PLB}{183}{91}{1987}.

\bibitem{RK}
G. Ripka and S. Kahana, \Journal{\PRD}{36}{1233}{1987}.

\bibitem{DP6}
D. Diakonov and V. Petrov, {\it Rotating Chiral Solitons Lie on Linear
Regge Trajectories}, preprint LNPI-1394 (1988) (unpublished);\\
D. Diakonov, {\it Acta Phys. Pol.} {\bf B25}, 17 (1994).


\bibitem{W2}
E. Witten, \Journal{\NPB}{223}{433}{1983}.

\bibitem{Bal}
A.P. Balachandran, A. Barducci, F. Lizzi, V.G.J. Rodgers and A. Stern,
\Journal{\PRL}{52}{887}{1984}.

\bibitem{DPPP}
D. Diakonov, V. Petrov, P. Pobylitsa and M. Prasza\l owicz,
\Journal{\PRD}{39}{3509}{1989}.

\bibitem{DPP3}
D. Diakonov, V. Petrov and M. Praszalowicz, \Journal{\NPB}{323}{53}
{1989}.

\bibitem{DJP}
D. Diakonov, J. Jaenicke and M. Polyakov, {\it Gluon Exchange
Corrections to the Nucleon Mass in the Chiral Theory}, preprint
LNPI-1738 (1991) (unpublished).

\bibitem{Review}
C. Christov, A. Blotz, H.-C. Kim, P. Pobylitsa, T. Watabe, Th. Meissner,
E. Ruiz Arriola and K.~Goeke, {\it Prog. Part. Nucl. Phys.}
{\bf 37}, 91 (1996), {\tt hep-ph/9604441}.

\bibitem{BR}
B. Golli, W. Broniowski and G. Ripka, \Journal{\PLB}{437}{24}{1998},
{\tt hep-ph/9807261};\\
G. Ripka and B. Golli, {\tt hep-ph/9910479}.

\bibitem{B}
A. Blotz, D. Diakonov, K. Goeke, N.W. Park, V. Petrov and P.~Pobylitsa,
\Journal{\PLB}{287}{29}{1992};
\Journal{\NPA}{355}{765}{1993}.

\bibitem{Gua}
E. Guadagnini, \Journal{\NPB}{236}{137}{1984}.

\bibitem{MNP}
P.O. Mazur, M.A. Nowak and M. Prasza\l owicz, \Journal{\PLB}{147}{137}
{1984}.

\bibitem{Chemtob}
M. Chemtob, \Journal{\NPB}{256}{600}{1985}.

\bibitem{JW}
S. Jain and S.R. Wadia, \Journal{\NPB}{258}{713}{1985}.

\bibitem{DPPol}
D. Diakonov, V. Petrov and M. Polyakov, \Journal{\ZPA}{359}{305}
{1997}, {\tt hep-ph/9703373}.

\bibitem{D3}
D. Diakonov, {\it Sov. J. Nucl. Phys.} {\bf 45}, 987 (1987)
[{\it Yad. Fiz.} {\bf 45}, 1592 (1987)].

\bibitem{DM}
D. Diakonov and A. Mirlin, {\it Sov. J. Nucl. Phys.} {\bf 47}, 421
(1988) [{\it Yad. Fiz.} {\bf 47}, 662 (1988)].

\bibitem{BlR}
J.-P. Blaizot and G. Ripka, \Journal{\PRD}{38}{1556}{1988}.

\bibitem{Dorey}
N. Dorey, J. Hughes and M. Mattis, \Journal{\PRD}{50}{5816}{1994}.

\bibitem{S}
P. Sieber, Th. Meissner, F. Gr\"ummer and K. Goeke,
\Journal{\NPA}{547}{459}{1992}.

\bibitem{WT}
T. Watabe and H. Toki, {\it Prog. Theor. Phys.} {\bf 87}, 651 (1992).

\bibitem{WY}
M. Wakamatsu and H. Yoshiki, \Journal{\NPA}{524}{561}{1991}.

\bibitem{WW}
M. Wakamatsu and T. Watabe, \Journal{\PLB}{312}{184}{1993}.

\bibitem{CGPPWW}
C. Christov, K. Goeke, V. Petrov, P. Pobylitsa, M. Wakamatsu and
T. Watabe, \Journal{\PLB}{235}{467}{1994}.

\bibitem{BPG}
A. Blotz, M. Prasza\l owicz and K. Goeke,
\Journal{\PRD}{53}{484}{1996}.

\bibitem{F}
R.P. Feynman, {\it Photon--Hadron Interactions} (Benjamin, 1972).

\bibitem{SF1}  
D. Diakonov, V. Petrov, P. Pobylitsa, M. Polyakov and C. Weiss,
\Journal{\NPB}{480}{341}{1996}, {\tt hep-ph/9606314}.

\bibitem{SF2}  
D. Diakonov, V. Petrov, P. Pobylitsa, M. Polyakov and C. Weiss,
\Journal{\PRD}{56}{4069}{1997}, {\tt hep-ph/9703420}.

\bibitem{PPner}
P.V. Pobylitsa and M.V. Polyakov, {\tt hep-ph/0004094}.

\bibitem{GR1}  
M. Gl\"uck, E. Reya, and A. Vogt, \Journal{\ZPC}{67}{433}{1995}.

\bibitem{GR2}  
M. Gl\"uck, E. Reya, M. Stratmann and W. Vogelsang,
\Journal{\PRD}{53}{4775}{1996}.

\bibitem{flavor2}  
P.V. Pobylitsa, M.V. Polyakov, K. Goeke, T. Watabe, and C. Weiss,
\Journal{\PRD}{59}{034024}{1999}, {\tt hep-ph/9804436}.

\bibitem{flavor}   
B. Dressler, K. Goeke, M.V. Polyakov, C. Weiss,
{\it Eur. Phys. J.} C {\bf 14}, 147 (2000), {\tt hep-ph/9909541};
{\tt hep-ph/9910464}.

\bibitem{PP}   
P. Pobylitsa and M. Polyakov, \Journal{\PLB}{389}{350}{1996},
{\tt hep-ph/9608434}.

\bibitem{E866Peng}  
J.C. Peng {\it et al.} (FNAL E866/NuSea Collaboration),
{\tt hep-ph/9804288}.

\bibitem{Bartels}  
J. Bartels and M. Loewe, \Journal{\ZPC}{12}{263}{1982}.

\bibitem{AFS} H. Abramowicz, L. Frankfurt, and M. Strikman,
in {\it Proceedings of SLAC 1994 Summer School} (SLAC Report 484),
p.539; {\it Survey High Energy Physics} {\bf 11}, 51 (1997),
{\tt hep-ph/9503437}.

\bibitem{Ji1}  X. Ji,
\Journal{\PRL}{78}{610}{1997}; {\it ibid.} {\bf D 55}, 7114 (1997).

\bibitem{Rad}
A.V. Radyushkin, \Journal{\PLB}{380}{417}{1996}.

\bibitem{CFS}
J. Collins, L. Frankfurt and M. Strikman,
\Journal{\PRD}{56}{2982}{1997}.

\bibitem{Rad2}   
A.V. Radyushkin, \Journal{\PLB}{385}{333}{1996}.

\bibitem{Piller}
L. Mankiewicz, G. Piller and T. Weigl,
\Journal{\PLB}{425}{186}{1998}

\bibitem{Ji3}  X. Ji, W. Melnitchouk and X. Song,
\Journal{\PRD}{56}{1}{1997}.

\bibitem{PPPBGW}
V. Petrov, P. Pobylitsa, M. Polyakov, I. B\"ornig, K. Goeke
and C. Weiss, \Journal{\PRD}{57}{4325}{1998}, {\tt hep-ph/9710270}.

\bibitem{PePG} 
M. Penttinen, M.V. Polyakov and K.Goeke,
\Journal{\PRD}{62}{014024}{2000}.

\bibitem{strikfurtmax}
L.L. Frankfurt, P.V. Pobylitsa, M.V. Polyakov and M. Strikman,
\Journal{\PRD}{60}{014010}{1999};\\
L.L. Frankfurt, M.V. Polyakov, M. Strikman and M. Vanderhaeghen,
\Journal{\PRL}{84}{2589}{2000}.

\bibitem{MPR}   
L. Mankiewicz, G. Piller and A.V. Radyushkin, {\it Eur. Phys. J}
C {\bf 10}, 307 (1999).

\bibitem{kroll}   
J. Boltz, P. Kroll, \Journal{\ZPA}{356}{327}{1996}.

\bibitem{MYNWF}
V. Petrov, M. Polyakov, P. Pobylitsa (in preparation).

\bibitem{pionwf}
V. Petrov, M. Polyakov, R. Ruskov, C. Weiss and K. Goeke,
\Journal{\PRD}{59}{114018}{1999}.

\bibitem{maxchrist}
M. Polyakov and C. Weiss, \Journal{\PRD}{59}{091502}{1999}.

\bibitem{COZ}
V.L. Chernyak, A.A. Ogloblin, I.R. Zhitnitsky,
\Journal{\ZPC}{42}{569}{1989}; {\it ibid.} {\bf 42}, 583 (1989).

\end{thebibliography}
\end{document}